\documentclass[12pt,preprint]{aastex}
\newcommand{\cc}{\mbox{cm$^{-3}$}}

\def\m17{M~17}               
\def\Htwo{H$_2$}               

\def\h{^{\rm h}}

\def\m{\ts {\rm m}}
\newcommand\kms{\rm{\, km \, s^{-1}}}

\def\swash2o{$1_{10} - 1_{01}$}             

\let\ts=\thinspace

\newcommand\etal{{\rm et al}.\ }
\newcommand\be{\begin{equation}}
\newcommand\en{\end{equation}}
\def\h2{\rm H_2}
\def\simgreat{^{>}_{\sim}}

\begin{document}

\title{Molecular Cloud Formation Behind Shock Waves}

\author{Edwin A. Bergin}
\affil{University of Michigan, 825 Dennison Building, 501 E. University Ave.,
Ann Arbor, MI 48109-1090} \email{ebergin@umich.edu}
\author{Lee W. Hartmann, John C. Raymond}
\affil{Harvard-Smithsonian Center for Astrophysics, 60 Garden Street,
Cambridge, MA 02138} \email{lhartmann@cfa.harvard.edu,
jraymond@cfa.harvard.edu}

\author{Javier Ballesteros-Paredes} 
\affil{
Centro de Radioastronom\'\i a y Astrof\'\i sica,
UNAM. Apdo. Postal 72-3 (Xangari), Morelia, Michoc\'an 58089, M\'exico
} \email{j.ballesteros@astrosmo.unam.mx}

\slugcomment{accepted by The Astrophysical Journal}
\begin{abstract}

Motivated by our previous paper, in which we argued for the formation of
molecular clouds from large-scale flows in the diffuse galactic interstellar medium,
we examine the formation of molecular gas behind shocks in atomic
gas using a one-dimensional chemical/dynamical model.  In our analysis we place
particular emphasis on constraints placed on the dynamical 
evolution by the chemistry.
The most important result of this study is to stress the importance of 
shielding the molecular gas from the destructive effects of UV radiation.
For shock ram pressures comparable to or exceeding typical local 
interstellar medium pressures, self-shielding controls the formation time 
of molecular hydrogen but CO formation requires shielding 
of the interstellar radiation field by dust grains.   
We find that for typical parameters the molecular hydrogen fractional abundance
can become significant well before CO forms.
The timescale for (CO) molecular cloud formation is not 
set by the H$_2$ formation rate on grains, but rather by the timescale
for accumulating a sufficient column density or extinction, $A_V \gtrsim 0.7$.

The local ratio of atomic to molecular gas (4:1), coupled with short estimates for
the lifetimes of molecular clouds (3-5 Myr), suggests that the timescales for
accumulating molecular clouds from atomic material typically must be no longer 
than about 12-20 Myr.  Based on the shielding requirement, this implies that
the typical product of pre-shock density and velocity must be 
$n v \gtrsim 20\;\cc\ {\rm km s^{-1}}$.  In turn, depending upon the shock velocity,
this implies shock ram pressures which are a few times the typical estimated
local turbulent gas pressure, and comparable to the total pressures (gas plus magnetic
plus cosmic rays).  Coupled with the rapid formation of CO once shielding is sufficient,
flow-driven formation of molecular clouds in the local interstellar
medium can occur sufficiently rapidly to account for observations.

We also provide detailed predictions of atomic and molecular emission and
absorption that track the formation of a molecular cloud from a purely
atomic medium, with a view toward helping to verify cloud formation by shock waves.
However, our predictions suggest that the detection of the pre-CO stages will be 
challenging.  Finally, we provide an analytic solution for time-dependent 
$\h2$ formation which may be
of use in numerical hydrodynamic calculations.
\end{abstract}


\section{Introduction}

The formation of stars and planetary systems is one of the fundamental
problems in astrophysics.   Much of the work over the past decades
has examined the formation of low mass stellar systems because these objects
sometimes form in isolation and are therefore easier to 
study individually.  One of the earliest stages of stellar birth that 
has been the focus of numerous investigations is the creation of a centrally
concentrated molecular core from a portion of
a low-density parent giant molecular cloud (GMC).  
Theoretical models account for the condensation as occurring possibly
via the slow diffusion of magnetic flux occurring over long
timescales ($\sim 10$ Myr) (Mouschovias 1999; Lizano \& Shu 1989) or the dissipation 
of turbulence on shorter timescales (Stone, Ostriker, \& Gammie 1998; 
Mac Low \etal 1998; Myers \& Lazarian 1998; Nakano 1998).  
Observations of isolated pre-stellar molecular
cores, such as L1544 in the Taurus Molecular Cloud, have provided
fertile ground for comparison to these theories (Caselli et al 2002; 
Ciolek \& Basu 2000; Williams et al 1999; Tafalla et al 1998).  However,
it is now recognized that most stars form in groups -- from small
aggregates to large clusters -- and it is not clear that all theories
developed for isolated star formation are easily applicable to the 
larger scales and simultaneity required for star cluster formation 
(Ballesteros-Paredes, Hartmann, \& V\'azquez-Semadeni 1999).

Expanding to larger scales opens the question as to whether the formation
of stars, both isolated and clustered, might perhaps be intimately related 
to the formation of the GMC itself.   
In a previous paper (Hartmann, Ballesteros-Paredes, \& Bergin 2001 
$=$ HBB01) we pointed out that the great majority
of molecular cloud complexes in the solar neighborhood appear to
be forming young stars, and that the ages of the stellar populations
in these clouds are typically $\sim$~2~Myr; stellar associations of
ages $\simgreat$~10 Myr are devoid of molecular gas.
The calculations indicating that MHD turbulence damps
rapidly (Stone et al. 1998, Mac Low et al. 1998, Padoan \& Nordlund 1999)
also favor the suggestion of short cloud lifetimes, since there is
no need for a continuous regeneration of MHD turbulence; additional 
support is found through arguments related to 
to cloud crossing times (Elmegreen 2000; HBB01).
These results place significant empirical constraints 
on the mechanism(s) of nearby molecular cloud formation
\citep[see][for a review]{maclow_rvmp, balles_cform, elmegreen_araa}.
For example, it is difficult to reconcile for the
transient nature of local clouds with models in which 
complexes are built up by the coalescence of smaller molecular
clouds, or theories in which molecular gas is mostly moved around
from one place to another (see Elmegreen 1993 and references therein),
because these processes are likely to take much more than $\sim 10-20$ Myr
to occur.   Moreover, short cloud lifetimes also place severe constraints 
on the processes leading toward fragmentation of the GMC and the condensation 
towards star formation.  

In HBB01 we suggested that chemical transformations of local gas are
essential to understanding the observational constraints.  We suggested
that most clouds are formed by large scale flows in the diffuse atomic
medium, and that they appear as molecular clouds only when the
column density becomes high enough to shield the molecular gas from the
dissociating effects of the interstellar radiation field (ISRF).
We further noted that self-gravity is likely to become important for
column densities comparable to that needed for shielding (also Franco
\& Cox 1986), which would explain the rapid onset of star formation
after molecular cloud formation.  Finally, we suggested that dispersal
of star-forming gas is accompanied by a reduction in shielding, so that
the gas may revert to an atomic state some time before it is completely
physically removed from the neighborhood.  

These suggested chemical transformations lessen, but do not necessarily
eliminate, the difficulty of making clouds ``fast enough''.  The formation
of H$_2$ from atomic gas, which generally must precede the formation of CO, is not 
instantaneous.  In addition, while
H$_2$ formation places important constraints on the problem,
it must also be examined in the
context of a model that incorporates the effects of H$_2$ (and CO) self-shielding from
the ultraviolet (UV) radiation field with 
extinction by dust grains, each with its own associated timescale. 
Following the molecular evolution is critical to an understanding of
cloud formation because of observational bias; molecular clouds are
essentially defined not by H$_2$ emission but through CO emission. 
In addition, possible atomic precursors are difficult
to identify against the galactic H I background (Ballesteros-Paredes \etal 1999).

In this paper we explore the formation of molecular gas in plane-parallel
shocks, starting with atomic (neutral) (warm) gas.  
Our model incorporates all the relevant heating and cooling mechanisms appropriate
for the interstellar medium (ISM), 
including chemical processes relevant
to the transformation of atomic gas to molecular form.
Koyama \& Inutsuka (2000) considered a similar problem, and did include the
important effects of shielding from UV radiation, but only examined 
maximum column densities of standard H~I clouds ($\sim 10^{19} - 10^{20}$ \cc ). 
Thus, they only achieve molecular hydrogen fractions of a few percent, 
insufficient to follow cloud formation.

We explore a range of shock velocities comparable to 
the flows expected in the diffuse atomic medium 
(Ballesteros-Paredes, Hartmann, \& V\'azquez-Semadeni 1999,
and references therein); the parameter ranges are also
appropriate for models in which GMC formation is induced by galactic spiral
density waves or energetic supernova.   Moreover, by following the 
primary cooling lines in the atomic shock and the post-shock evolution we
can predict fluxes
for key transitions of C II, C I, O I, CO and other species
spanning a range of initial ram pressures.  This should aid
in the search for the progenitors of molecular clouds.  Finally as a by-product
of this work we provide an analytic solution for the formation of 
molecules behind a shock and discuss the relation between gas temperature 
and extinction, both of which may be of some use to MHD modeling.

In \S 2 we describe the model with results and a parameter study
provided in \S 3. 
In \S 4 we outline the observational possibilities for the detection
of forming molecular clouds via both emission and absorption lines.
Section 5 summarizes the implications of these results for star formation
in the local neighborhood and in \S 6 we present our conclusions.

\section{Model}

We combine the results of two separate one-dimensional plane-parallel 
models to examine the full thermal and dynamical evolution of parcel of 
atomic gas that is heated by a shock, condenses, and cools via radiation while 
slowly evolving to molecular form.   The first model (labeled as
the {\em Atomic} code) examines the initial
shock wave propagating through a purely atomic medium.   This code, described
in \citet{raymond79} and \citet{coxray85},  has been extensively used to 
examine the coupled dynamical and ionization evolution of shocked gas in the ISM.   
The second model (labeled as the {\em Molecular} code) is a time dependent 
chemical model that has been used to examine molecular chemical evolution 
in the dense regions of the galaxy that are currently forming stars \citep{bl97,bmn98}.
This model computes the time evolution through the atomic/molecular transition 
and has been modified to include the same flow equations, dominant 
heating and cooling processes as \citet{raymond79} \citep[but see also][]{cox72}.  
Additional processes more relevant to the atomic/molecular transition have
also been included.  These will be discussed below.

\subsection{Dynamical Solution}

In each code the fluid equations for a steady state plane-parallel flow
are used \citep{cox72}:  

\begin{equation} 
nv = n_0 v_0
\end{equation}

\begin{equation} 
P + \rho v^2 + \frac{B^2}{8\pi} = n_0(1 + f_0/1.16)kT_0 + \rho_0 v_0^2 + \frac{B_0^2}{8\pi}
\end{equation}

\begin{equation} 
\frac{dh}{dt} = -\frac{\Lambda}{n} + \frac{\Gamma}{n}
\end{equation}

\noindent where $n$ is the density in cm$^{-3}$, $v$ is the gas velocity,
$\rho$ is the mass density, $P$ is the pressure, and $B$ is the magnetic field.
$h$ is the enthalpy per particle, while $\Lambda$ and $\Gamma$ are the 
cooling and heating rates respectively.   These equations are solved
simultaneously with the chemistry to provide atomic and molecular abundances, 
gas temperature, and gas density as a function of distance/time.  
In the following we assume that $ n = n_{\rm H I} + 2n_{\rm H_2}$, but in our 
dynamical calculations we have accounted for the presence of heavy atoms.

In the one-dimensional framework as the shock propagates through 
the medium there is a build-up of colder and denser material behind the shock.
Gas and dust in this column will absorb interstellar UV radiation.   The majority of 
absorption is by dust grains, but as molecules begin to form self-shielding
of H$_2$ or CO molecules can become important.   The visual extinction can
be estimated via the time integration of equation (1):

\begin{equation}  
2 \times 10^{21} A_V = \int_{0}^{t} nv dt = \int_{0}^{t} n_0 v_0 dt = n_0v_0t.
\end{equation}  

\noindent in this equation $t = 0$ is assumed to be the moment of the initial shock
in the medium.   This calculation has been included in both models to examine
the attenuation of X-rays and the UV photons responsible for gas 
heating and photodissociation.

\subsection{Initial Conditions and Chemistry}

In our models we have assumed that a shock, with speeds up to 
50 $\rm km~s^{-1}$, propagates into 
the galactic warm neutral medium with an initial temperature of 1000~K.  The
pre-shock density is set to the average density of the ISM $n$ = 1 cm$^{-3}$,
but higher pre-shock densities have been explored.
In all the models examine a range of shock ram pressures from 1 -- 30 $\times
10^4$ cm$^{-3}$ K.
Since dust is required for the 
UV shielding we have adopted depleted metal abundances based on
observations of $\zeta$ Oph.   Abundances and references are provided
in Table~1.  Species are assumed singly ionized or neutral
depending upon whether their ionization potential is above
(He, N, O, Ne, Ar) or below (C, Mg, Si, S, Ca, Fe, Ni) 13.6 eV.
At the start of the calculation H is predominantly neutral with a fractional
ionization of 10$^{-2}$ in the form of H$^+$.

The choice of magnetic field strength necessitates special comment.
As noted by many authors, even a very small component of the
interstellar magnetic field tangential to the flow can prevent compression
sufficient to make molecular gas.  If the ram pressure of the shock is
equal to the typical value in the local ISM, the equipartition magnetic 
pressure occurs $B = 6 \mu$G; any field parallel to the shock of this
magnitude will prevent further compression.  In a plane-parallel steady
shock, the parallel field will be linearly proportional to the density,
$B \propto n$.  If we wish to compress the gas from an initial density
of, say, 3 ${\rm cm^{-3}}$ to $\sim$300 ${\rm cm^{-3}}$, the parallel magnetic field
must increase by a factor of at least 100 from its initial value.  But $6 \mu$G is the
typical rms field strength in the diffuse ISM.  Thus, unless the initial field
parallel to the shock is only $\lesssim$ 1\% of the initial total field strength,
compression to form a cloud cannot occur.  This qualitative argument is not changed
unless the ram pressure is enormously greater than the typical ISM value.

HBB01 suggested that the solution to this
problem is indicated by the MHD calculations of Passot, V\'azquez-Semadeni,
\& Poquet (1995); specifically, that clouds tend to form at bends or ``kinks''
in the magnetic field, allowing material to flow along the field and finally
compress in a direction perpendicular to the magnetic field.  (See Figures 4 and
5 of HBB01 and related explanation in the text.)  Putting it another way; flows
which are {\em not} preferentially along field lines (bent or not)
{\em cannot} form even diffuse clouds, let alone molecular clouds.
Thus clouds must preferentially occur when conditions are such that the magnetic
field does not prevent compression; flows in other directions, for other
conditions, simply don't make clouds.  Based on this argument, and the results
from the simulations of Passot \etal (1995),
we assume that the {\em initial} magnetic field component parallel to the shock
front is not dynamically important, adopting a nominal value below 0.01 $\mu$G.
Note that the component of the magnetic field perpendicular to the shock front
is unconstrained in this one-dimensional model.  Further, we note that even
with such a small initial parallel magnetic field, the post-shock parallel
magnetic field strength is not negligible in some cases, and can approach
$10 \mu$G.   

The  {\em Atomic} and {\em Molecular} models include relevant chemical
processes specific for the treated phase.    
The {\em Atomic} model self-consistently solves for the time-dependent ionization
state of all species included in Table~1.
However, this model does not include molecular processes.  In particular, 
the H$^-$ mechanism to form H$_2$ is not incorporated, as this process is
more important in 
fast shocks with significantly higher 
pre-shock densities and greater ion fraction \citep{hm79, nd89}.   
This model evolves until the post-shock gas temperature reaches 1000~K at which
point the {\em Molecular} model adopts the final physical and chemical 
state from the {\em Atomic} code as initial conditions.
All important parameters (density, temperature, abundances, Ly $\alpha$ flux,
heating rates, cooling
rates, ionization fraction) are uniform to within a few percent at this hand-off.

The {\em Molecular} model self-consistently solves for the time dependent chemistry 
of molecular formation from a purely atomic medium 
using a minimized chemical network that focuses on the formation
of H$_2$, CO, and other simple carbon and oxygen-bearing molecules.
The system of ordinary differential equations governing the chemical evolution
is solved using the DVODE algorithm \citep{vode}. 
In this calculation a conservation equation is included for each 
atomic pool. 
The minimized rate network has been used previously to examine the formation of water
and molecular oxygen in dense molecular clouds \citep{bmn98}, and under those
conditions has been tested extensively through successful comparisons
to a larger network \citep{umist}.   

The model has been adapted to include the formation of H$_2$ via surface
recombination on grains with a rate provided in Table~2.
We have assumed that the dust temperature is low enough to allow for
H$_2$ formation at the rate estimated in the ISM by Jura (1974).
The calculations of Cazaux \& Tielens (2004; see also
Cazaux \& Tielens 2002) indicate a near-unity efficiency of formation
of H$_2$, provided the dust temperature does not exceed $\sim 20$ K.  
In the ISM the average dust temperature derived from COBE data lies between
16 -- 23 K (Reach et al. 1995); one expects that in regions with the needed
dust extinction for shielding (see below), prior to star formation, the dust
temperature will lie closer to the lower value.

The {\em Molecular} model does not treat the ionization state of atomic species beyond
the singly ionized phase.  This assumption is valid as most highly ionized atoms 
recombine early in the post-shock phase that is treated by the 
{\em Atomic} model prior the hand-off.
One additional simplification in the {\em Molecular} model is that it does not include 
all the atomic species listed in Table~1.  Instead 
the most abundant atomic species are included (H, He, C, O) 
with atomic and neutral iron as the sole representative heavy ion.   
This difference does not affect our results.

For the photodissociation rates
of molecular species we use the rates compiled by \citet{umist}, except
for CO and H$_2$ where we adopt the self-shielding formalisms of
\citet{lee_coshield} and \citet{draine_h2shield}.
These approximations include the effects of dust absorption and 
molecular line shielding.
The H$_2$ self-shielding approximation (Eq. 36) of \citet{draine_h2shield} is well
matched to more exact models with a turbulent velocity parameter of 3 km s$^{-1}$.  The purely
thermal line width in our calculations is generally well below this value.
We do not correct for this difference, and under the assumption of small turbulent
widths, our models would under-predict the effects of H$_2$ shielding.  
This is not a question for CO, because the dissociation lines are 
broadened due to pre-dissociation \citep{ah99}.
We have also assumed the radiation field from the shock is not
greater than the ISRF.  Any significant ultraviolet radiation from
the shock would lengthen timescales for molecular formation. 
However, our
favored solutions involve weaker shocks ($v \le 10$ km s$^{-1}$)
impinging upon gas with
above average density.  Such shocks would produce smaller amounts
of UV radiation that will not significantly alter our primary results. 
Even the faster shocks we consider produce
little UV radiation aside from Ly~$\alpha$,
and the large H~I column in the atomic
part of the flow effectively shields the
molecular region from Ly~$\alpha$ photons.

We note that the {\em Atomic} model does not work well for shock speeds below $10\;{\rm km s^{-1}}$.
However, for many purposes the {\em Molecular} model behavior is independent of the
results from the {\em Atomic} model, as long as the same pressure is maintained.  
Thus, one may extrapolate many of the results of the {\em Molecular} model to slower shocks
with the same ram pressure.

\subsection{Heating, Cooling, and Thermal Balance }

Both the {\em Atomic} and {\em Molecular} models include the relevant
heating and cooling processes for each regime.  Thus both codes contain
photoelectric, X-ray, and cosmic ray heating.   The {\em Molecular} model
also includes heating due to the formation of H$_2$.  Each model
contains cooling via C II, C, and O line emission
and H$^+$ recombination.  The molecular model also includes cooling
via CO rotational emission in the ground vibrational state. 
Following the example of 
\citet{wolfire95} in  Table~2 we summarize the major
thermal processes and references included in these models.   

In the initial stages of the post-shock evolution, while the excess
dynamical energy is lost through radiation, the temperature and
compression factor ($x = n/n_0$) are determined 
using the fluid equations (\S 2.1) and methods outlined by Cox (1972).   
At later stages when the dynamical energy of the shock has dissipated the 
temperature is derived through thermal balance of the heating and cooling
terms via standard Newton-Rhapson techniques (Press et al 1988).  Pressure
balance then provides the compression factor. 
These solutions are determined in tandem with the chemical evolution.

\section{Results}

Our standard model assumes an initial shock velocity of $v_0 = 20$ km s$^{-1}$
impacting gas with a temperature of $T_0 = 1000$~K and a density of $n_0 = 1$
\cc ($\rho v^2 =$ 28000 cm$^{-3}$ K).  This shock speed is consistent with that induced by either spiral density waves 
\citep{shu_spiral, roberts_spiral} or turbulence driven collisions
between flows (Ballesteros-Paredes et al 1999).   
A supernova will likely induce a series of shocks
with a range of ram pressures covered by this standard solution 
and the additional 
runs that explore parameter space.

In all calculations presented below the {\em Atomic} model is used to examine the 
gas evolution while the temperature exceeds 1000~K.  When gas temperature
in the post-shock evolution decays to below this value the {\em Molecular} model
is used to examine the subsequent evolution.

\subsection{Standard Model}

\begin{figure}
\includegraphics[height=20cm]{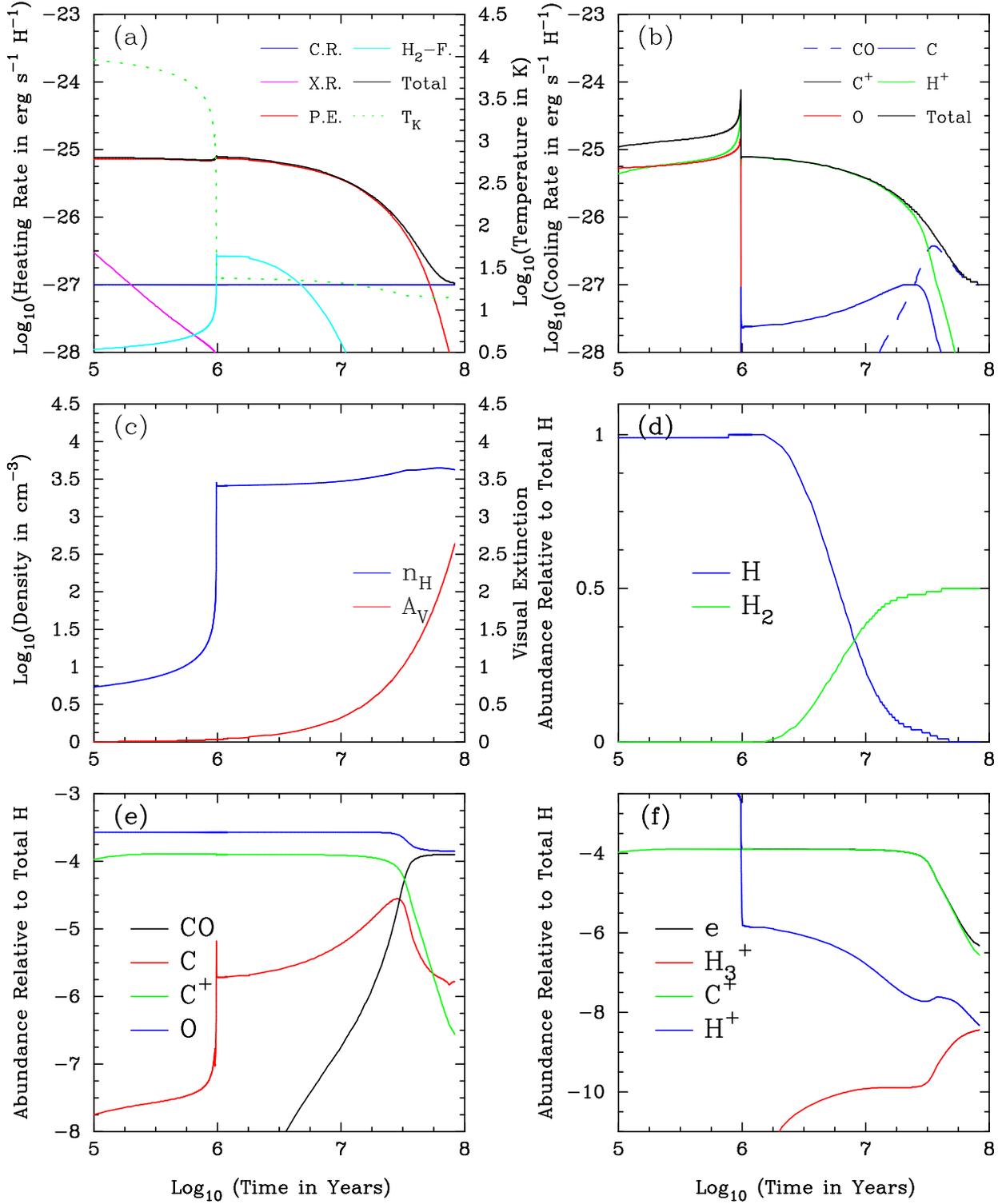}
\caption{Figure showing the time evolution of relevant quantities
for a model assuming $v_s = 20 \kms$, $n_0$ = 1 \cc , and $T_0$ = 1000 K.
In the following panel (a) shows the heating rates and temperature 
(shown as dotted line with axis labeled to right), (b) 
cooling rates, (c) density 
and extinction with labels to left and right respectively, (d) 
H and H$_2$ abundances, (e) carbon and oxygen abundances, 
(f) major electron contributors. }
\label{basic}
\end{figure}

Figure~\ref{basic}(a-f) presents the physical and chemical evolution of the post-shock
gas.  Here panels (a-c) show the major heating and cooling terms along with the
primary physical parameters (density, temperature, optical extinction) as a 
function of time.  Panels (e-f) show the chemical abundances (relative
to total H) of important molecular and atomic species as a function of  time.
In this figure t = 0 represents the onset of the shock.

After the shock the gas temperature is quite high ($\sim 10,000$ K) with 
the primary cooling is via [C II], [O I], [Si II], and [Fe II]
line emission (hydrogen line emission is also an important coolant 
in the initial stages).  At early 
times ($< 10^6$ yr) the heating processes 
which dominate the cold neutral medium (CNM),
photoelectric and X-ray heating, are unimportant.  
At t = 10$^6$ yrs the temperature
drops sharply at the cooling timescale set by the overall cooling rate.
For this case ($v_s = 20 \kms $ and an
initial density of 1 cm$^{-3}$) the cooling timescale 
is $(3/2)kT/\Lambda \sim 10^6$ yr (Spitzer 1978).

When the gas cools the density rises due to pressure equilibrium.
At this point the solution reaches a stable high density
atomic phase where 
photoelectric heating is balanced by 
[C II] cooling.  This stable ``plateau'' is essentially the 
Wolfire et al. (1995) CNM solution and is roughly
at the temperature of the CNM as measured by \citet{ht03}.  
The plateau exists for $\sim 3 \times 10^6$~yr until the onset of
H$_2$ formation which is followed by the formation
of CO.  At this point CO begins to dominate the cooling reducing 
the temperature
to $\sim 10$ K.  Photoelectric heating dominates throughout most
of the post shock evolution until A$_V > 0.3$~mag, whereupon cosmic-ray
heating is more important.\footnote{At these 
low densities the gas is thermally decoupled
from dust grains.}   Prior to the formation of CO there is a 
period  where the abundance of neutral carbon rises and it contributes
to the cooling, but it is not 
the dominant carbon reservoir.  

The timescale of this process is controlled by the formation
time of H$_2$ on grains and the slow buildup of shielding due to grains
and H$_2$ molecules downstream in the post-shock gas.    
At the density of the stable plateau (n $= 2500$ cm$^{-3}$; T = 23 K) 
the H$_2$ formation time is $\sim 4 \times 10^6$ yr.   The onset of molecular
formation occurs at much later times and therefore {\it the evolution is controlled 
by the shielding of UV radiation.}
In our standard case the shielding of H$_2$ is dominated by self-shielding
with only small contributions from dust absorption.  For CO the
shielding is dominated through the UV photon absorption by dust grains. 

\subsection{Parameter Search}

In the following we will examine different initial ram
pressures through changes in the shock speed given the same
initial density and temperature as the standard model.  
We also examine a few cases with a constant shock velocity but varying 
initial densities, 
to verify the similarity of results for similar
ram pressures for the molecular gas.

\begin{figure}
\includegraphics[width=6cm, angle=-90]{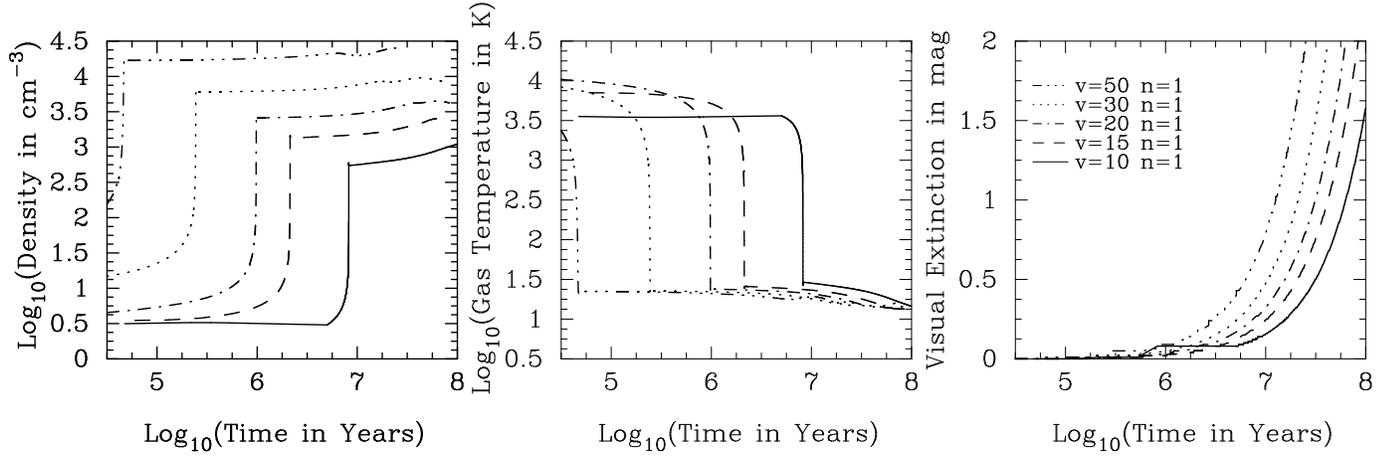}
\caption{Temporal evolution of hydrogen density (left), gas 
temperature (middle), and extinction in the visual (right) for 
models with variable
shock strength labeled to right.  Each model adopts $n_0$ = 1 \cc\
and $T_0$ = 1000 K.}
\label{phys_ev}
\end{figure}

\begin{figure}
\includegraphics[height=16cm]{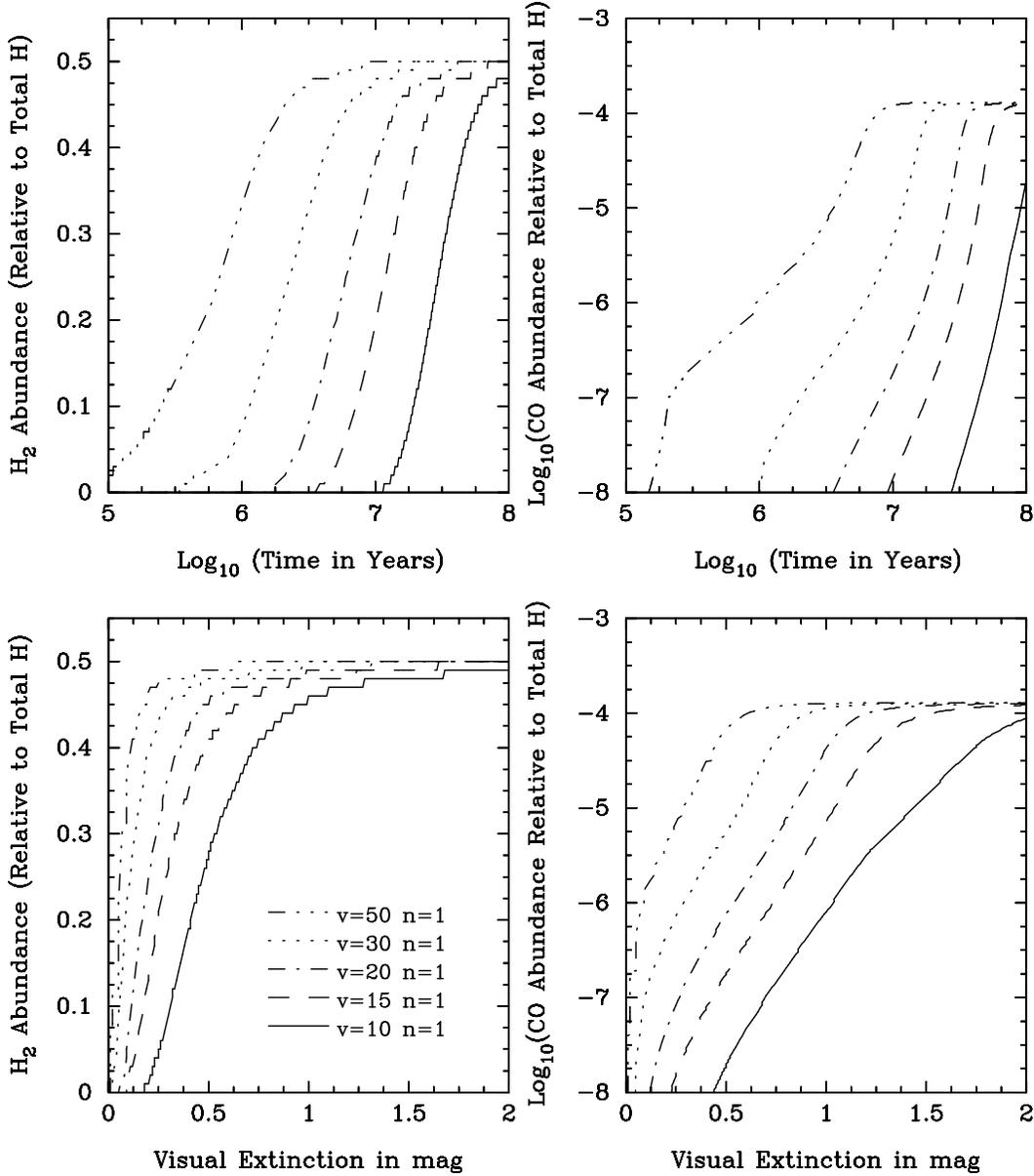}
\caption{Top panels show the time evolution of the H$_2$
(top left) and CO (top right) abundance.
Bottom panels show the evolution of the H$_2$ abundance 
(bottom left) and CO abundance (bottom right) as a function of
visual extinction.   All abundances are relative to total H. 
Each panel shows models with different shock speeds which are labeled
in the bottom left panel.  Each model adopts $n_0$ = 1 \cc\ 
and $T_0$ = 1000 K.  }
\label{abun_ev}
\end{figure}

In Figure~\ref{phys_ev} we present the results from models with 
$v_s = 10, 15, 20, 30, 50 \kms$ ($T_0 = 1000$~K; $n_0 = 1$ \cc ),
which have ram pressures ranging from 1.4 -- 36 $\times 10^4$ cm$^{-3}$ K. 
The H$_2$ and CO abundance as a function of time and visual extinction
are given in Figure~\ref{abun_ev}.
Rather than show the detailed evolution (as in Figure~\ref{basic}) 
we present the temporal evolution of salient parameters: density,
temperature, and extinction.  
We note that the cooling time to T $< 10,000$ K is 
extremely short due to collisional excitation of H I.  Thus
the much higher immediate post-shock temperatures in the faster
shock models do not appear in Figure~\ref{phys_ev}.

In Figure~\ref{phys_ev} the cooling timescale to reach the stable temperature 
plateau decreases with increasing ram pressure.   Similarly the
timescales of molecular formation decrease (Figure~\ref{abun_ev}).   
Both effects are due to the density of the solution plateau increasing with
ram pressure.
For H$_2$, in each case, except for the 10 $\kms$ shock, 
self-shielding dominates over dust absorption.  The reverse
is true for CO, dust absorption dominates over self-shielding, except
for a 50 $\kms$ shock, where self-shielding becomes more important
at A$_V > 0.5$~mag. 
 
\begin{figure}
\includegraphics[width=6cm, angle=-90]{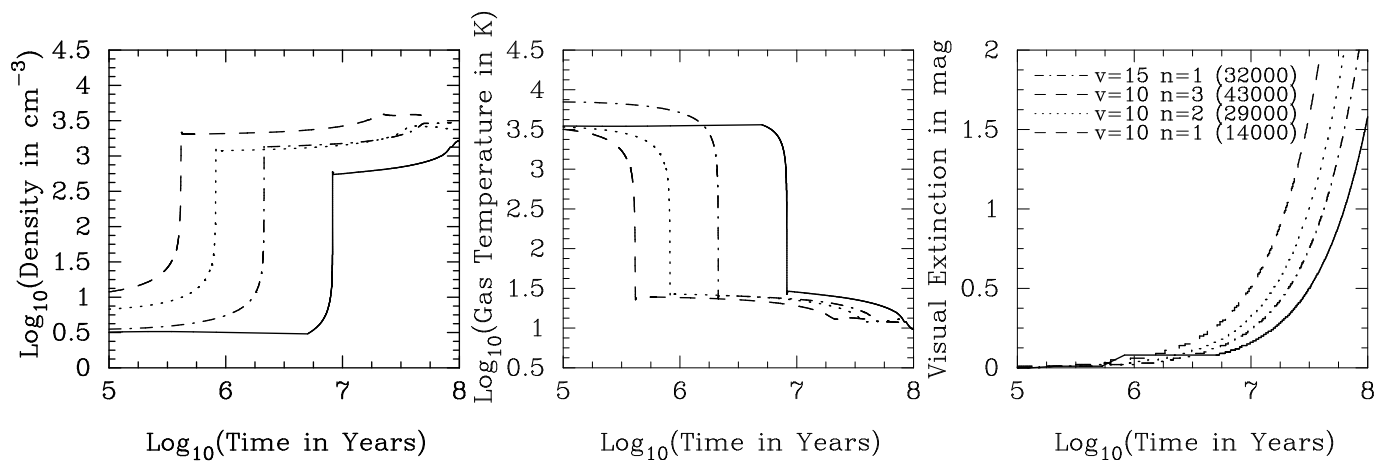}
\caption{
Illustration of the effects of ram pressure and mass conservation.
Temporal evolution of hydrogen density (left), gas 
temperature (middle), and extinction in the visual (right) for 
models with variable
density labeled to right.  
The density in cm$^{-3}$ and initial shock velocity in K
are provided in the bottom left panel.  The shock ram pressure
in cm$^{-3}$ K is given in parenthesis.
}
\label{phys_ev_ic}
\end{figure}

\begin{figure}
\includegraphics[height=16cm]{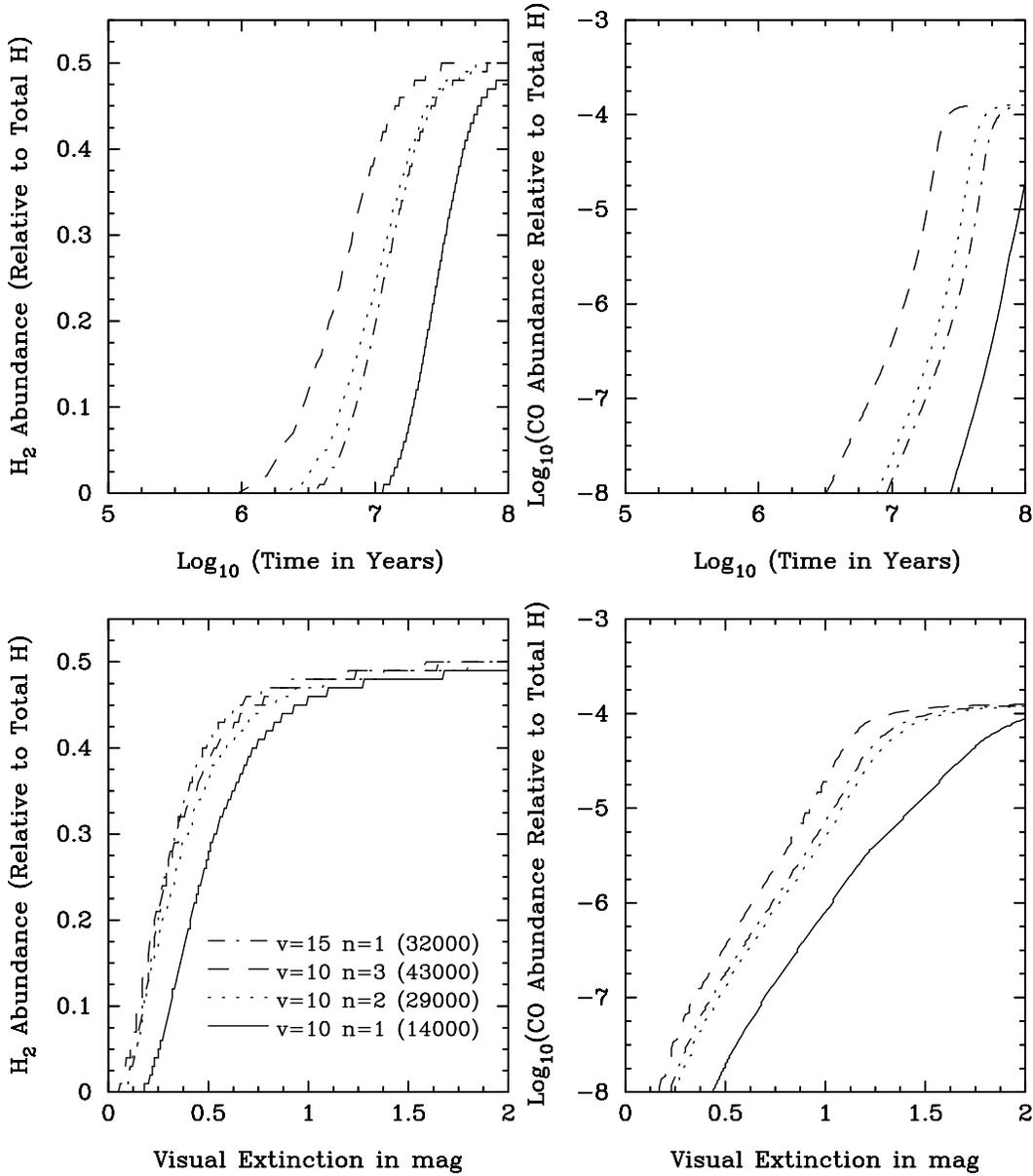}
\caption{Illustration of the effects of ram pressure and mass conservation.
Top panels show the time evolution of the H$_2$
(top left) and CO (top right) abundance.
Bottom panels show the evolution of the H$_2$ abundance 
(bottom left) and CO abundance (bottom right) as a function of
visual extinction.   All abundances are relative to total H. 
The density in cm$^{-3}$ and initial shock velocity in K
are provided in the bottom left panel.  The shock ram pressure
in cm$^{-3}$ K is given in parenthesis.
}
\label{abun_ev_ic}
\end{figure}

In Figures~\ref{phys_ev_ic} and \ref{abun_ev_ic} we examine models with the same shock
speed ($v_s = 10 \kms$) but different initial densities.
We also show the $v_s = 15 \kms$,
$n_0 = 1$ \cc\ case as it has similar ram pressure 
as the $v_s = 10 \kms$, $n_0 = 2$ \cc\ solution. 
Here we see similar effects as illustrated by models with increasing
shock velocity, all timescales (cooling, molecular formation) shorten
with increasing density.
Models with similar ram pressure (compare  
P/k = 32000~cm$^{-3}$~K with P/k = 29000~cm$^{-3}$~K) 
have slightly different evolution in the physical and chemical
properties seen in Fig. 4 and 5.   Indeed the model with the
slightly smaller ram pressure has faster physical and chemical evolution.
In this model the temperature decay occurs more quickly because of 
the higher initial density, which increases the cooling
through collisionally excited lines.
Subsequently, H$_2$ forms earlier, and the higher $nv$ allows for 
faster CO formation.   Overall the differences between these two cases
are not large and are magnified by the log scale.
However, this points out that the effects of mass conservation 
and pressure conservation are not equivalent.   
Solutions with higher densities will have
faster dynamical evolution through increased cooling and quicker
{\em molecular} evolution due to the decrease in the time to
reach full shielding.   

\section{Observational Evidence for Molecular Cloud Formation}

In principle the primary observational evidence for the cloud formation 
process would be H~I emission from dense ($n > 1000$~cm$^{-3}$)
purely atomic gas.
However, in practice this is difficult because the excitation
characteristics of 21~cm emission, and the pervasive presence of warm
H~I emission in the galactic plane, makes it difficult to disentangle 
cold and warm gas along a given line of sight
\citep{spitzer78, kulkarni_heiles}.  Some cold 
H~I clouds have been isolated from emission studies
\citep{knee_brunt}, but their relation to forming molecular clouds
is unclear.  A better way to detect cold atomic hydrogen is via
absorption against warm H~I background emission; and this has
received more attention in recent years \citep{gibson_hi, li_hi}

In general, the search for molecular cloud formation has typically 
concentrated on the cold stages of the evolution (i.e. not the shock
itself but on the build up of molecular material).  
Thus catching a cloud undergoing
incipient formation can be pursued by searching for gas that is dense
by galactic ISM standards ($n \gg 1$ cm$^{-3}$), with emission from atomic
carbon but not associated with CO emission.
However, any mechanism (e.g. spiral density wave, supernova shock, 
turbulent flow
collision in ISM)  that produces a shock wave which compresses the gas
will produce emission and absorption
associated with both warm (T $>$ 1000~K)  and cold (T $= 10 - 40$~K) gas.  
Thus the important cold phase can be searched for via  traditional
tracers (e.g. [C I], CO), but traces of the shock may also
be evident in atomic emission and absorption. 

In the following sections we will present our model results which 
provide detailed predictions of atomic and molecular emission and
absorption 
tracing the formation of a dense condensation of molecular gas.

\subsection{Line Emission}
\subsubsection{Optical lines}

Table~3 shows the relative line intensities of the stronger atomic emission
lines.  They are generally similar to the spectra of slow shocks used to
interpret Herbig-Haro object spectra (Hartigan, Morse \& Raymond 1994),
except for the effects of the much higher densities in the HH object shocks.
Depending on shock speed, the spectral signature may be strongly
enhanced H$\alpha$ or strongly enhanced [S II] emission.

The surface brightness of a slow shock in tenuous gas is quite low,
typically a fraction of a Rayleigh, and a shock seen face-on would be
difficult to detect.  However, such a shock seen edge-on would be
geometrically brightened in the same manner as emission filaments
in supernova remnants.  Edge-on shocks might be found near star forming
regions, but depending on the thickness of the colliding clouds, the
shock emission may cease before stars are formed.  Therefore, the
emission might be best seen as an extension of a filamentary molecular
cloud.

The shocks studied here might also be detected through their cumulative
contribution to the diffuse Galactic emission.  For instance, Haffner et al
(1999) report that some heating beyond that supplied by photoionization is
required to account for the strengths of [N II] and [S II] lines detected
at low spatial resolution by the Wisconsin H$\alpha$ Mapper (WHAM).
Minter \& Spangler (1997) suggest that dissipation of turbulence might
provide that extra heating.  If that dissipation occurs in shocks, the
shocks in the Warm Ionized Medium (WIM) would provide the [N II], while
those in the Warm Neutral Medium (WNM) would contribute [O~I] and [S II].

\subsubsection{Far-Infrared and (sub)-Millimeter lines}

\begin{figure}
\includegraphics[height=6cm]{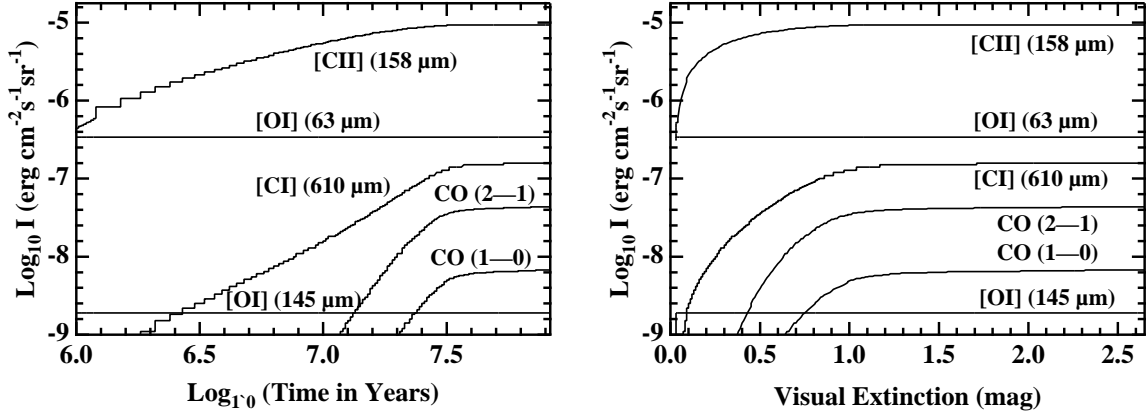}
\caption{
Predicted line intensities (erg cm$^{-2}$ s$^{-1}$ sr$^{-1}$) for selected
atomic and molecular transitions arising from the plane-parallel slab
as a function of time (left-panel)
and visual extinction (right-panel) in the standard model.
For reference 1 K km s$^{-1}$ is 1.22 $\times 10^{-7}$ for the
[C I] 610 $\mu$m line, 1.6 $\times 10^{-9}$ for CO (1-0), and 
1.2 $\times 10^{-8}$ for CO (2-1).
}
\label{intensity}
\end{figure}

While much of the shock energy is released via cooling lines in the 
UV, optical, and infrared,  the majority of emission in 
cool post-shock phase lies in far-infrared and millimeter wavelengths.
In Figure~\ref{intensity} we provide a plot of line intensity 
(erg cm$^{-2}$ s$^{-1}$ sr$^{-1}$) as a function of time and 
visual extinction for our standard model ($n = 1$ \cc\ and $v_0 = 20$
$\kms$).  Figure~\ref{intensity} shows only the evolution during the cold
post-shock phase, but  includes contributions to emission from
the atomic shock model.  In this Figure some transitions
(e.g. [O~I]) do not emit during the cold post-shock evolution, while
others ([C~II], [C~I], CO) show a slow steady increase in emission
with time due to the build up of column behind the 
shock.\footnote{For space considerations we provide only one sample
of our results for a specific set of conditions.   
As seen in \S 3 we have examined a wide range of parameter space
and additional plots can be provided upon request to the first author. }

\begin{figure}
\includegraphics[height=6cm]{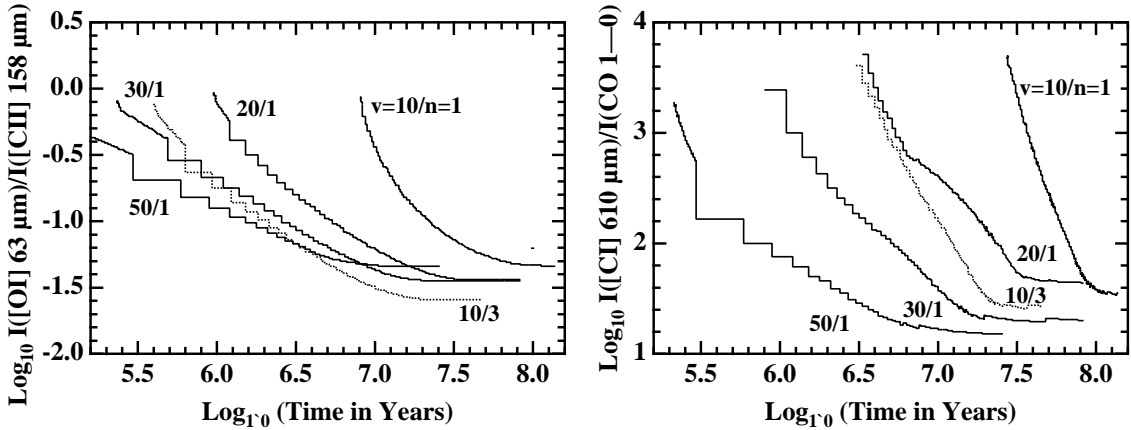}
\caption{
Left: Ratio of the [O I] 63 $\mu$m 
to the [C II] 158 $\mu$m line intensity shown as a function of time.
Right: Ratio of the intensity of the [C I] 610 $\mu$m line to the 
CO (1-0) line.   
In both panels several models are shown with a key
of velocity/density of the model.   Thus 20/1 is a model with a shock
velocity of 20 km s$^{-1}$ and an initial density of 1 \cc .
}
\label{Iratio}
\end{figure}

In Figure~\ref{Iratio} we provide diagnostic plots of line intensity ratios for
specific transitions that can be used as suitable tracers of 
cloud formation.  In this plot examples are given for a range
of parameter space.   Because there is little [O~I] emission in the cold
phase the line ratio of [O~I](63~$\mu$m)/[C~II](158~$\mu$m) exhibits
a factor of $\sim$30 drop until it saturates when CO has formed and
removes carbon ions from the gas.  The intensity ratio of
[C~I](610~$\mu$m)/CO~(1-0) also strongly varies (by nearly 2 orders of
magnitude) over much of the evolution until the ratio again flattens
due to the formation of CO.   A key question for these diagnostics
is whether this signature is unique or is indistinguishable with galactic
photodissociation region (PDR) emission arising from  gas with similar conditions.
To examine this question we use the PDR models of \citet{kaufman_pdr}.

In the PDR model the [O~I](63~$\mu$m)/[C~II](158~$\mu$m) ratio ranges
from 0.03 $-$ 0.3 for conditions that mimic those in our models
(n $\sim 1000 - 10^4$~\cc ; G$_0 = 1.7$).  This range is quite similar
to that seen in our results; only for a short time at the start of the 
post-shock evolution is the ratio significantly greater.   Thus 
this ratio is not a strong signpost of molecular cloud formation.
However, the [C~I](610~$\mu$m)/CO~(1-0) ratio does differ from
PDR models (which predict ratios between 25 -- 50 for similar
conditions).  This is essentially due to the increased column of
atomic carbon during stages prior to CO formation, in PDR models the 
atomic carbon is created via photodissociation of CO.   However, it should
be noted that the [C~I](610~$\mu$m) 
emission is quite weak
at times where the [C~I](610~$\mu$m)/CO~(1-0) ratio
is highly elevated above typical PDR values. 
Best results appear to require
deep searches for [C~I](610~$\mu$m)
emission to a level below $\sim$0.1 K km s$^{-1}$ 
(I = 1.22 $\times 10^{-8}$ erg cm$^{-2}$ s$^{-1}$ sr$^{-1}$).

\subsection{Absorption Lines}

\subsubsection{Total Column Densities}

In some instances favorable placement may provide a background source
to detect cold post-shock gas absorption.  For 
H~I, galactic hydrogen emission provides a nearly ubiquitous background.
In our models the total column for a given species is determined
by the initial shock conditions, the chemistry, and also by 
the time evolution.  Thus, there is no single value of the column density.
In Table~4 we provide a sample of the predicted column
densities of atomic and molecular tracers as a function of shock model
and for a sampling of total extinction (A$_v$ = 0.2, 0.5, and 1.0~mag). 
For atomic hydrogen we have divided the column by noting the amount
that resides in warm (T $\gg 60$ K) and cold (T $< 60$ K) 
gas.\footnote{The column of gas between 60 -- 1000~K is quite small and is included in the
warm column.}

For the weakest shock ($v_s = 10$ $\kms$, $n = 1$ \cc ) the 
warm H~I column is significant in comparison to the column of cold gas.  
However, the cold gas column dominates in all other cases.  Although
the warm gas column is not provided for O~I and C~II, similar results 
are seen for these atoms as well. 
Typically the warm column is $\sim$few 
\% or less of the column at any given extinction.  
In the case of CO and C~I, these species only form when the gas
has cooled.  

Some general trends are seen in Table~4. First, the warm and cold
H~I column is lower for stronger shocks, while  
C~I and CO columns increase with shock velocity/ram pressure.
This is due to the faster
chemical and physical timescales at the higher post-shock densities.
In contrast because neutral oxygen and ionized carbon are the 
respective atomic reservoirs during much of the evolution, their 
column densities are
relatively insensitive to shock strength, showing only a factor of $< 3$ 
variation at the highest extinction.  

\subsubsection{Atomic Carbon Fine-Structure Excitation}

\begin{figure}
\includegraphics[height=9cm]{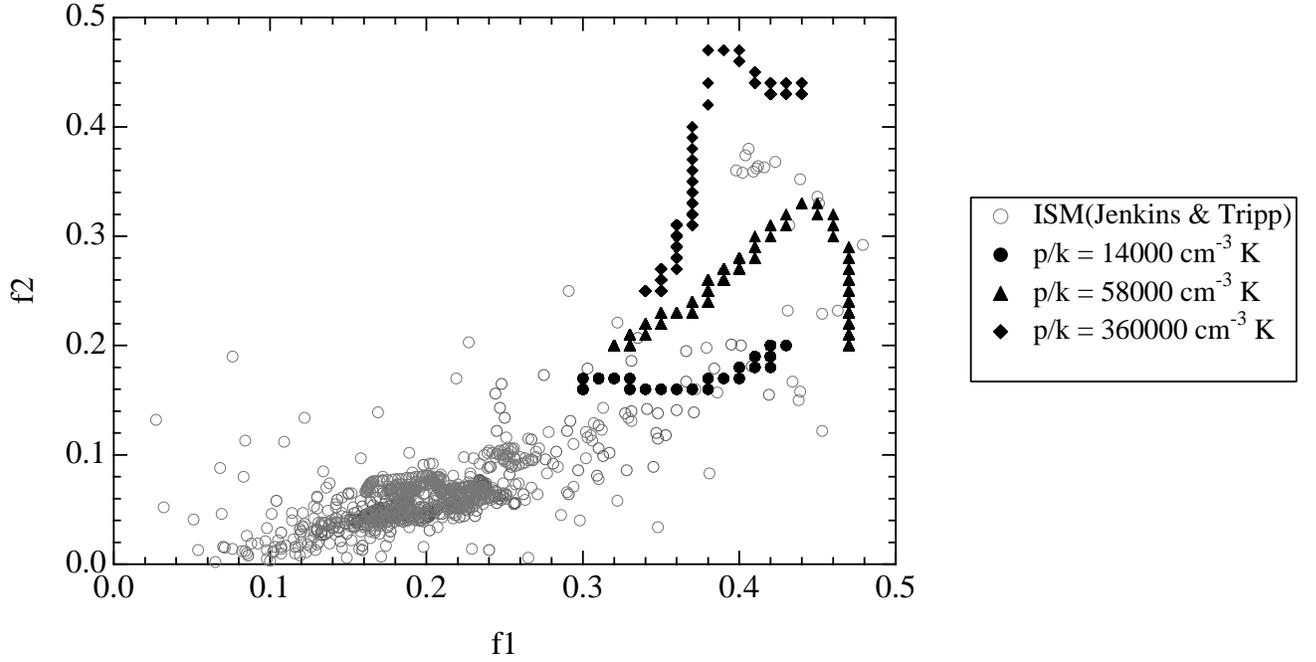}
\caption{
Plot of the fractional population of the 1st (f1 = $^3P_1$) and
2nd (f2 = $^3P_2$) carbon atom fine structure levels.  The open grey circles
are observed values in the interstellar medium taken from the observations
of Jenkins \& Tripp (2001).  The solid markers denote the range of values
predicted in our models with variable shock strength (ram pressure 
provided in key) and
with an initial density of 1 \cc .
The plotted ratios are the column density of gas in each of the
fine-structure levels at a given time ($^3$P$_2$ and $^3$P$_1$) over
the total column and thus accurately reflect what would be observed.
The range in values for given shock parameters is due to the time evolution
of the post-shock gas. 
}
\label{carbon}
\end{figure}

\citet[][hereafter JT01]{jenkins_tripp} used UV 
absorption observations of atomic
carbon multiplets to determine the relative populations in fine
structure levels along numerous galactic lines of sight.   
Based on excitation ratios they find that the 
median pressure in the CNM 
is $\sim p/k = 2240$ cm$^{-3}$ K and that 
many lines of sight exceed this median value.  In Figure~\ref{carbon}
the grey points show the distribution of the 1st atomic carbon 
fine structure level fractional population (f1 = $^3P_1$) as a
function of the 2nd (f2 = $^3P_2$).
These numbers have been derived along numerous lines
of sight in the ISM by JT01.  
The cluster of points
between f1 = 0.12 -- 0.25 and f2 = 0.0 -- 0.1 show typical CNM ratios
near the median pressure.  Based on excitation analyses,
clouds with f1 $>$ 0.3 and f2 $>$ 0.1 receive some contribution
from over-pressurized gas (JT01).\footnote{ 
Absorption lines arising gas with these ratios typically have
velocities outside those permitted by galactic rotation.} 

\citet{jenkins_tripp} theorized that the
pressure enhancements are the result of compression due to converging
turbulent flows or warm boundary layers in the surfaces of dense clouds
moving within a lower density medium.  
The predicted fine-structure excitation ratios from our models, 
which are of the shock creation of 
such over-pressurized gas, are also shown in Figure~\ref{carbon}.
We have selected three models, each with an initial density of $n = 1$ 
cm$^{-3}$ and shock velocities of 10, 20, and 50 $\kms $, with corresponding
to ram pressures of 1.4, 5.8, and 36 $\times 10^{4}$~cm$^{-3}$~K.
The predicted ratios overlap with the ``high'' pressure gas 
seen by JT01, which demonstrates that high C I excitation ratios 
are an additional
signature of the shock.  Weaker shocks, which evolve on longer timescales,
are likely to be more prevalent in the ISM 
and thus these shocks  ($P/k < 1.4 \times 10^4$~cm$^{-3}$~K)
appear more representative of observations.   
We did not model such weak shocks as they will not evolve on the 
timescales required for fast star formation (see \S 5.1).
However, the excellent agreement of our models with the high excitation
end of this distribution suggests that the JT01 results could indeed
be the result of a distribution of weak and strong shocks 
active in the atomic medium throughout the galaxy.

\section{Discussion}

\subsection{Molecular cloud formation by atomic flows}

Our model calculations raise an obvious question: when should we consider
that a ``molecular cloud'' has been formed?  As shown in Figures 1 and 3,
$\h2$ can form at considerably earlier times and lower column densities
than CO, especially for low ram pressures/shock velocities.  Thus low-pressure
clouds could become molecular, in terms of the dominant constituent, before
they become CO clouds.  However, $\h2$ is difficult to detect, and is not
usually a criterion for defining a molecular cloud.  We consider the problem
of $\h2$ detection in the next subsection; here we concentrate on the detection of CO.

It is useful to distinguish between what we call the ``accumulation''
timescale, the length of time it takes to accumulate $A_V \sim 0.7$ from the
atomic gas and the cloud becomes ``detectable'' in CO, 
from other timescales, for instance the evolution of the CO abundance.
For purposes of discussion we have made the somewhat arbitrary decision that the
cloud appears when CO J=1--0 emission can be detected at a level of
1 K km/s -- which defines the accumulation timescale.

\begin{figure}
\includegraphics[height=6cm]{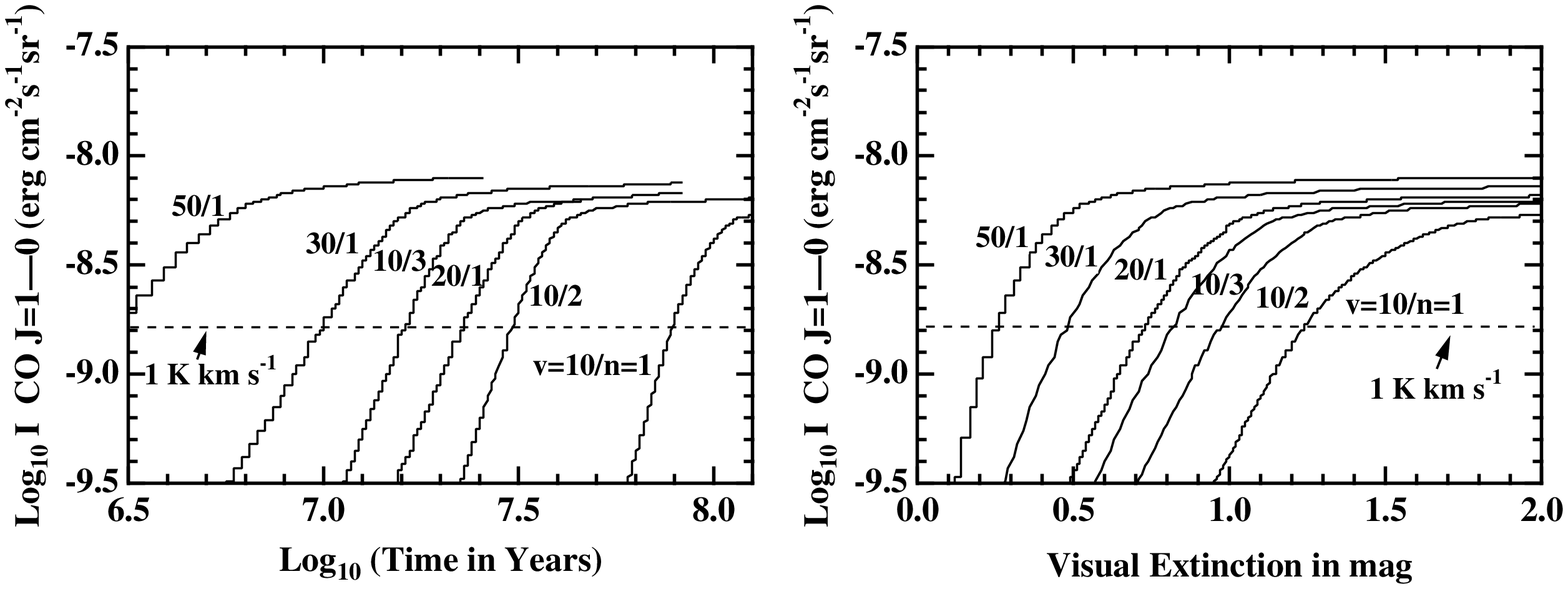}
\caption{
Predicted line intensities (erg cm$^{-2}$ s$^{-1}$ sr$^{-1}$) for CO J=1--0 
arising from the plane-parallel slab
as a function of time.  Several models are shown with a key
of velocity/density of the model.   Thus 20/1 is a model with a shock
velocity of 20 km s$^{-1}$ and an initial density of 1 \cc .
The dashed horizontal line denotes a CO J=1--0 integrated intensity of
1 K km s$^{-1}$.
}
\label{co_flux}
\end{figure}

In Figure~\ref{co_flux} we provide the intensity of the J=1--0 transition
of CO as a function of time with the dashed line denoting an integrated intensity
of 1 K km/s.  Based on this definition, the cloud accumulation 
timescale ranges from $\sim 3 \times 10^{6}$ yr for a 50 $\kms$ shock to  
$8 \times 10^7$ yr for a 10 $\kms$ shock (both with an initial density of 1 \cc ).
As noted before, we find that the column density or extinction is the most important
parameter in determining the formation of CO.  
Based on the right-hand panel of Figure~\ref{co_flux},
the above definition is roughly equivalent
to molecular cloud ``formation'' at $A_V \sim 0.7$ mag for ram pressures 
between $\sim 3 - 5 \times 10^4$ cm$^{-3}$ K.\footnote{We 
have assumed that the post-shock gas is not exposed to enhanced
radiation fields during any time of this evolution. If the shock front
passes close to a massive star than the evolution will be slowed
and the A$_V$ threshold will increase by an amount of $\sim ln(G_0(t)/4.4)$.
Here we list the UV enhancement factor G$_0$ as a function of time to denote
the fact that effects of radiation will diminish as the shock front moves
away from the star.}

The above threshold of $A_V \sim 0.7$ mag is based on a 1-dimensional calculation
and it is useful to discuss how this might change in a more realistic 3-dimensional
geometry.  Chemically the transition from CII/CI/CO will be seen at a level where
the CO photodissociation becomes ineffective.   In 1-D this threshold is found at 0.7 mag
with little contribution from self-shielding.   
In a three-dimensional calculation for a clumpy cloud, there will be two, competing,
effects: first, UV radiation will propagate more freely into the cloud; and second,
clumping may increase the local column densities and thus promote the transition
to molecular gas.  Our 1-D requirement of $A_V \sim 0.7$ would thus translate into
an average over solid angle of the extinction of the diffuse ultraviolet radiation
field.  The time to reach this threshold is no
longer a simple calculation of the $n*v*t$ product,  though we expect that
the transition to molecular gas will occur roughly when the average density/column density
reaches values comparable to those in our calculations.

We also note that in addition to clumping, overall contraction of the cloud
in the direction perpendicular to the shock front will promote rapid molecular
gas formation by increasing the density and the shielding.  Recent numerical
simulations of finite self-gravitating sheets (Burkert, A., \& Hartmann, L. 2004,
in preparation) suggest that global gravitational collapse of flattened
clouds is likely and rapid.

In HBB01 we showed that the stellar population ages of nearby molecular clouds imply cloud
lifetimes of no more than about 3-5 Myr.  Generally, slightly older associations are not 
immediately next to current sites of star formation, indicating that molecular material 
is not simply being pushed around but that there is some cycling between the atomic and 
molecular states.  The ratio of atomic to molecular gas in the solar neighborhood is 
estimated to be $\sim 4$ \citep{savage_h2, dame93}.  This implies that the timescale for turning 
atomic gas into molecular material, on average, is about $12 - 20$~Myr.  
(If only a fraction of the atomic gas cycles through molecular stages, the timescale for the
conversion of atomic to molecular gas in specified regions must be shorter.)
In our terminology the cloud accumulation timescale must therefore
be $<$ 20 Myr; the cloud lifetime would then be defined by the time from when
a cloud is detectable in CO emission until the gas is dissipated by star 
formation.

Thus, the one-dimensional constraint on the accumulation timescale timescale $t(acc)$ implied
by the ratio of atomic to molecular gas, coupled with the requirement that $A_V \sim 0.7$ 
for CO formation, constrains the possible average pre-shock parameters, 
\begin{equation}
n v \gtrsim 20 \, (20 Myr/t(acc)) \, \cc\ \kms \,.
\end{equation}
Can this constraint reasonably be satisfied?  The one-dimensional rms turbulent velocity 
of cold H~I in the local interstellar medium is $\sim 6 \kms$ (Boulares \& Cox 1990);
thus it seems reasonable to take a typical shock velocity $v \sim 10 \kms$.
Accumulation timescales of 10-20 Myr then require typical pre-shock densities $\sim 2-4\;\cc$.
This is a few times larger than the average density of the ISM, but it would not be particularly
surprising if molecular clouds were preferentially formed from initially slightly higher
densities.  
Moreover, it has been increasingly accepted that
the density fluctuations in the ISM (clouds) are produced primarily by
compressions due to a supersonic turbulent velocity field (Von
Weiszacker 1951; Sasao 1973; Elmegreen 1993; Padoan 1995; Ballesteros-Paredes,
V\'azquez-Semadeni \& Scalo 1999). In such an environment, the
production of the density fluctuations are likely the result of a
succession of compression events, so that denser structures are formed
by compressions within previously compressed, larger ones, rather than
a single, very strong one (V\'azquez-Semadeni 1994).  
This scenario naturally explains
the density probability density function observed in numerical simulations of isothermal and
polytropic flows \citep{pv98}, and here it naturally provides the necessary
conditions (2--4 cm$^{-3}$) for the pre-shock gas in our
calculations.
As already pointed out, ram pressure is the most important
parameter for CO formation; the ram pressures for the above parameters are a few times the
average turbulent gas pressure (gas plus magnetic plus cosmic rays)
in the ISM (Boulares \& Cox 1990); again, this does not
seem to be an unreasonable constraint for making the highest density regions in the solar
neighborhood. 

Once the shielding column density is achieved, CO formation is rapid.  Additional increases
in density due to subsidiary shocks and/or gravitational contraction will yield to even
more rapid CO formation.  At a column density corresponding to $A_V \sim 1$, comparable
to the shielding length estimated here, the characteristic
growth time for gravitational contraction in (subsonic) gas at $T \sim 10$~K is of the order    
of 1 Myr.  Thus, once molecular gas is formed, star formation can ensue rapidly,
at least in regions where supersonic turbulence has been dissipated (HBB01).

We conclude that our results are consistent with the picture presented in HBB01 of rapid
formation of molecular clouds from atomic material, as long as the starting densities are
typically a few times the average interstellar density.

\subsection{Molecular Hydrogen in the Atomic Medium}

Here we consider whether the presence of some molecular hydrogen in the pre-shock
gas would shorten either the accumulation timescale or the timescale for CO abundances
to rise. Indeed, because H$_2$ does not emit for
typical conditions in the atomic CNM it might be possible to hide a significant
molecular component.   

To examine this question we have examined solutions with a substantial
H$_2$ fraction in the pre-shock gas.  This is only performed using 
the {\em Molecular} model because the {\em Atomic} model does not include 
chemical processes linked to H$_2$.  
One limitation is that the cooling via H$_2$ emission 
and shock dissociation of H$_2$ are not included.
However, the addition of H$_2$ cooling will only shorten the cooling time to 
reach the stable plateau solution.  As we will show below, the chemical
evolution (i.e. CO formation) is dominated by the shielding timescale which
depends primarily on the initial shock parameters.

In Figure 10 we present the H$_2$ and CO abundances as a function of
time and extinction in our standard model with an initial H$_2$ fraction
of 0.0, 0.125, and 0.25.  The sharp rise of the H$_2$ abundance at
t = 10$^6$ yr is due to the non-inclusion of H$_2$ in the {\em Atomic}
model.  At this time the abundance of CO shows a sharp spike which is 
due to rapid CO formation from the pre-existing H$_2$.  
However, CO molecules are quickly dissociated,
due to the lack of UV shielding.
What is striking in these plots
is that even an a priori presence of H$_2$ molecules --
a necessary requirement for CO formation -- has little effect on the CO chemical
evolution.  This effect is discussed in \S 3, because the CO
formation requires dust shielding the evolution cannot proceed until
sufficient dust column exists.  

This result is robust provided the pre-shock gas is 
exposed to the ISRF with A$_V = 0$~mag.
In Figure 11 we present solutions with higher initial extinction and
an initial H$_2$ fraction of $f = 0.25$.  The left-hand panels show the
evolution of the CO concentration, while the right-hand panels present
the CO J=1--0 emission, which we have used to define the cloud accumulation 
timescale ($\tau_{acc}$ = time where emission reaches a level
of 1 K $\kms$).  With this definition even pre-shock gas with A$_V = 0.5$~mag
will only accumulate on timescales a factor of two shorter than the model starting
with unshielded gas.
Moreover the differences in the emitted J=1--0 intensity between these solutions
are negligible unless $A_V \ge 1$~mag.
In sum, the presence of H$_2$ in the pre-shock gas would not hasten the timescales
for cloud formation, unless the gas is already significantly shielded
from UV radiation and CO is already effectively in existence.

Let us now place these results in an observational context.
Allen and co-workers have argued for a pervasive molecular component 
on the basic of the relative placement of dust lanes, radio continuum, and H~I 
emission in external spiral galaxies \citep{allen_h2, tilanus_h2, allen_nat_h2}.   
In the standard picture of cloud formation
H~I emission would appear in front of the spiral shock wave as
traced by the radio continuum emission.  Instead these authors found that 
the H~I emission is observed  downstream of the shock, which is interpreted 
as the result of photodissociation of H$_2$ by young stars.   This led \citet{pringle_h2}
to theorize that there may be a significant reservoir of molecular gas in the
low density CNM which would allow for fast cloud and star formation.  They argue that this 
inter-arm gas could be hidden by having the temperature colder than 10 K, 
producing only weak CO emission.   

However, these results must be placed in the context of studies of
H$_2$ in absorption and CO in emission in our own galaxy.
The most extensive initial study of H$_2$ in absorption
was performed by the Copernicus satellite and \citet{savage_h2} found
that the average H$_2$ fraction within 500 pc of the Sun is $<f> = 0.25$.   
More recent
FUSE observations have demonstrated that H$_2$ absorption is pervasive, even
in sight lines well out of the galactic plane towards background extragalactic sources 
\citep{shull_fuse}.
In addition, an analysis of  X-ray absorption spectra along similar lines of sight 
by \citet{arabadjis_h2} finds that 
the required X-ray absorption column exceeds the
H~I column estimated by 21 cm emission.  This excess is difficult to account for 
via other atomic components, but could be due to molecular gas.  The UV and X-ray
observations point to the potential ubiquitous presence of H$_2$ in
the galaxy with fractions
of H$_2$ towards these high galactic latitude lines of sight ranging
from $\sim$10\%  \citep{shull_fuse} to $>100$\% \citep{arabadjis_h2}
of the H$_2$ fraction estimated in the local solar neighborhood.

While there is important evidence for the presence of some H$_2$ in the low 
density CNM, the evidence for CO is less
substantial.  For instance, the surface density of molecular 
gas estimated by CO emission towards high galactic latitudes is 
only 1\% of the value derived in the local neighborhood.   This is well below that
seen for the H$_2$ fraction and hints that CO may not be fully tracing H$_2$ in the 
low density ISM.  This is not terribly surprising, because
the column density needed for equivalent CO self-shielding 
is three orders of magnitude larger than that needed for H$_2$ 
formation (Lee et al. 1996; Draine \& Bertoldi 1996).

Given the possibility of a significant present of H$_2$ in the ISM one might question
whether significant reservoir of H$_2$ molecular clouds exist that are ``inert'' in
the sense that they will never create CO.  There are two possibilities that can be examined
in this regard.  First the \citet{jenkins_tripp} results suggest an abundance of weak shocks are
active in the galactic ISM.  These shocks will evolve on slow times for the creation of both
H$_2$ and CO.  Thus in the $> 30$ Myr required to make H$_2$ there is an additional
$> 30$ Myr before CO creation (see Fig. 3).  For such a weak shock there will be 
a significant amount of time spent in the molecular but pre-CO phase.
The overall lifetime of such systems may be limited by the large distance that would
be transversed.  For example, a $\sim$ 5 km s$^{-1}$ shock will
travel a distance $\gtrsim$500 pc and therefore has an increased likelihood of passing by a massive
star that will certainly destroy the CO and perhaps H$_2$.
An alternate picture would emerge if the shock evolution were limited
in some fashion (perhaps by a strong tangential B-field).  This will
halt the density evolution but column will continue to build up behind the shock.  
Thus, the combination of ensuing extinction evolution, H$_2$ molecules, and cosmic ray ionization
eventually leads to CO formation. However, if the density evolution were stunted at
a values below $< 1000$ \cc\ then the H$_2$ phase would last longer in a similar fashion as the
weak shock case.  
However, the importance of such inert clouds might be limited by passing hot stellar wind
or supernova bubbles which could disrupt or evaporate them.
In sum we cannot discount the possibility that there 
might be a hidden reservior of H$_2$ molecular gas in the ISM.

In this picture with ubiquitous H$_2$, and little CO, 
cloud formation timescales will depend primarily on the timescale to accumulate 
a sufficient shielding column density, not the H$_2$ formation time.
Hence the solution is similar to that found in the traditional case where cloud formation
is modeled from the atomic-molecular hydrogen transition.

\subsection{Numerical Models of ISM Dynamical Evolution}

Complex MHD 
numerical simulations of structure formation in the ISM, such as those
performed by \citet{osg, bhv99, maclow98}, typically neglect the chemical
considerations when computing the evolution. This is easily understood given
the extreme computational complexities.   In our simulations we have included
the chemistry but with a simpler dynamical prescription.   In this process
we have identified two areas where this work can have useful impact on 
numerical simulations of cloud formation and evolution.

Our results point to the key importance of the shielding of UV photons 
in the overall physical evolution.   In Figure 12 we show the temperature
evolution as a function of the mass surface density (visual extinction).
The initial spikes below 1 g cm$^{-2}$ are the shock and initial cool down.
The following evolution shows the gradual temperature cool down as
the shielding column increases.  For all shocks the 
gas temperature, at a given surface density, is nearly identical,
to within a factor of $< 1.5$, and declines in a linear fashion in all cases
except for the strongest shock.  This difference is due to the lack
of atomic carbon cooling which typically contributes 
to the cooling at A$_V \sim$ 1 (in the fast shock quick CO formation via 
self-shielding reduces the influence of C~I).
The similarities of these solutions and the 
linear decay suggests that tracking the shielding in each cube of a
MHD simulation can be used to provide a simple estimate of the gas temperature
and hence pressure. 

Another potential aid to MHD simulations is provided in the appendix where
we show that there exists an analytical solution for the time-dependence
of H$_2$ formation (CO formation can be treated via equilibrium calculations).
This analytical solution requires only knowledge of the local gas density
and extinction. 
Under those conditions computationally
intensive chemical calculations need not be performed, while still keeping
the capability to create a realistic simulation that includes the
chemical formation of the two most important molecular species: H$_2$ and CO.
If this can adapted into MHD models then it provides 
a method to use simulations to predict maps of 
molecular emission, which can be readily compared to observations.

\section{Conclusions}

We have presented a detailed coupled physical and chemical model that examines
the  shock compression of atomic gas and the slow transition to molecular form.
This model
includes all of the physical and chemical processes relevant to investigate
the formation of molecular clouds via shocks induced by cloud-cloud collisions,
spiral density waves, or supernovae within the dynamic interstellar medium. 
Our principle results are as follows:

1) We find that the molecular cloud formation timescale is not controlled by
the formation rate of H$_2$ on grains.  Rather
the shielding of molecules from the UV radiation is the limiting 
parameter.    For all but the weakest shocks we find that H$_2$ self-shields quite
efficiently.  However,  CO formation requires shielding of the interstellar radiation
by dust grains.  Thus the cloud formation timescale is effectively set by 
the time needed to accumulate a column equivalent to A$_V \sim 1$ mag in extinction. 

2) Molecular cloud formation times can be as short as $\sim 10 - 20$ Myr,
as required by our picture of rapid cloud formation from large scale flows,
adopting typical velocities in the ISM ($v \sim 10 \kms$) for starting densities
and ram pressures a few times higher than average interstellar values.

3) Since shielding is required for CO formation the a priori presence of H$_2$
in the low density medium will not appreciably shorten the time required
to create molecular clouds.   

4) We provide detailed predictions of the atomic and molecular
emission and absorption that trace the formation of molecular clouds.  
A subset of these predictions match current conditions observed
in over-pressurized gas within the cold neutral medium of the galaxy by 
Jenkins \& Tripp (2001). 

5) A by-product of this work is an examination of ways to incorporate the effects
of chemistry into detailed MHD simulations of structure formation in the galaxy.
This includes an analytic solution for the time-dependent formation of
molecular hydrogen and a discussion of the overall temperature structure of
dense cooling proto-clouds.

\acknowledgements

We thank an anonymous referee for helpful suggestions which improved the
manuscript.  EAB is grateful to E. Jenkins for kindly providing 
his CI excitation data and for discussions with E. Ostriker and J. Bregman.
This work was supported in part by NASA grants NAG5-9670 and NAG5-13210.
This research has made use of NASA's Astrophysics Data System Service.

\appendix

\section{Time Dependent Analytical Solution for H$_2$ Formation}

For more complex applications it is useful to have an analytical approximation
to time dependence of the formation of $\h2$ , following a shock in the
low density gas.  A reasonable approximation can be derived through the
differential equation that describes the $\h2$ evolution:

\begin{equation} 
\eqnum{A1}
\frac{dn(\rm{H}_2)}{dt} = R_{gr}(T_p )n_p n(\rm{H}) - [\zeta_{cr} + 
\zeta_{diss}(\it{N}(\rm{H}_2),A_V)]\it{n}(\rm{H}_2).
\end{equation} 

\noindent Here $n_p$ is the total density, $T_p$ is the temperature,
$R_{gr}(T_p )$ is the temperature dependent
grain formation rate
of \Htwo\ in cm$^3$ s$^{-1}$, $\zeta_{cr}$ the \Htwo\ cosmic ray ionization rate, and
$\zeta_{diss}$ the \Htwo\ photodissociation rate, which is a function of the visual extinction
and the total \Htwo\ column density.  In equilibrium $dn$(\Htwo )/$dt = 0$ and
$n$(H)/$n$(\Htwo ) = ($\zeta_{cr}$ + $\zeta_{diss}$)/$R_{gr}(T_p )n_p$.  

Using the relation between the total density with hydrogen atoms and molecules
($n$(H) + 2$n$(\Htwo ) = $n_p$) Eqn. (A1) can be rewritten:

\begin{equation} 
\eqnum{A2}
\frac{dn(\rm{H}_2)}{dt} = R_{gr}(T_p )n_p^2 - [2R_{gr}(T_p )n_p + \zeta_{cr} + 
\zeta_{diss}(N(\rm{H}_2),A_V)]\it{n}(\rm{H}_2).
\end{equation} 

\noindent If we set $R_{gr} n_p^2$ $= Z$ and $(2Rn_p + \zeta_{cr} + \zeta_{diss}) = Y$ then

\begin{equation} 
\eqnum{A3}
\frac{dn(\rm{H}_2)}{dt} = Z - Yn(\rm{H}_2).
\end{equation} 

\noindent Strictly speaking the total density, \Htwo\ formation rate (temperature dependent), 
\Htwo\ photorate (through the column density build up) have time-dependencies.  
However, after the shock the evolution quickly converges to a
roughly constant plateau in density and temperature, which is essentially  
the Wolfire et al. (1995) solution for non-shielded ISM gas.  This plateau in the
physical evolution persists until shielding decreases the photo-electric heating rate
and allows CO formation.  This ultimately 
lowers the temperature and raises the density by factors of 2--3.
Thus, for a given shock pressure,
the density and temperature can be approximated using values from the
non-shielded ``Wolfire'' solution (here: $n_p$, $T_p$).  
The H$_2$ photo-rate can be treated in a different fashion, by discretizing 
Eqn. (A3) and lagging the build-up of the H$_2$ column (and the determination
of the photo-rate)  behind the H$_2$ density solution.   In this fashion the
H$_2$ photodissociation rate from the previous step is used in the current one.
These approximations both result in some
loss in accuracy, but as will be shown below the difference between the analytical
and exact treatments is not large. 
With these assumptions, Eqn. (A3) can therefore be solved,

\begin{equation} 
\eqnum{A4}
n(\rm{H_2}) = \it{Z/Y - Ce^{-Yt}},
\end{equation} 

\noindent where $C$ is a constant.  Using the $t$ = 0 boundary condition
($n$(H$_2) =$ 0), the constant can be determined and, $C = Z/Y$.   
Substituting back, $Y \equiv 1/t_0 =$ 2/$\tau_{dep}$ 
+ 1/$\tau_{cr}$ + 1/$\tau_{ph}$ and $C = n_p t_0/\tau_{dep}$:

\begin{equation} 
\eqnum{A5}
n(\rm{H_2}) = \it{n_p t_0/\tau_{dep}}(\rm{1} - \it{e^{-t/t_0}}).
\end{equation} 

\noindent  In this expression $\tau_{dep}$ is the depletion timescale,
$\tau_{cr}$ is the cosmic ray ionization timescale, and $\tau_{ph}$ is the
timescale for photodissociation including self-shielding and dust shielding.

This expression can be used to derive a reasonable estimate of the H$_2$ evolution, with
knowledge of the \Htwo\ photodissociation rate.
Using the approximation of Draine \& Bertoldi (1996), two factors are required 
to estimate the  H$_2$ photo-dissociation timescale:
(1) the dust extinction at 1000\AA\ and (2) the H$_2$ column density.  
(1) The extinction can be estimated though the mass conservation equation 
as described
in \S 2.1, using $\tau_{dust,1000} =$ $n_0 v_0 t \sigma_{dust,1000}$, where $\sigma_{dust,1000}
= 2 \times 10^{-22}$ cm$^{2}$ ($n_0$ and $v_0$  are
the initial pre-shock density and velocity, and $\tau_{dust,1000}$ is
the dust opacity at 1000 \AA ).  {\em In simulations the identification of
a shock may not be straightforward.  The most important quantity
is the tracking of the 
average 2-D/3-D total column, which can
be converted to the opacity using  $\sigma_{dust,1000}$.}
(2) As discussed above,
the H$_2$ column density can be approximated through the buildup of the column
behind the shock.

Figure 13 (left) shows the fraction of \Htwo\ as a function of time for the exact solution for
the 15 km s$^{-1}$ shock model along with the analytical solution.  We also show for  
comparison, the exact treatment of a limiting case where the \Htwo\ is fully shielded from
the start.  Here the analytical solution assumes $n_p = 1400$ cm$^{-3}$ and $T_p = 20$ K, which
is the evolutionary plateau for the 15 km s$^{-1}$ shock with a pre-shock density
of 1 cm$^{-3}$.
In this figure the analytical solution lies between the 
exact solution and the fully shielded calculation.  However, the agreement between
the analytical and exact treatments is 
within a factor of two for relevant timescales.   The analytical model
over-predicts the amount of \Htwo , because the calculation cannot mirror the full 
treatment of the density and temperature evolution, which leads to increased \Htwo\ 
self-shielding.   In the above example, the density at the early stages of the
full treatment is $\ll$1400 cm$^{-3}$, thus the analytical example will over-predict
the early creation of \Htwo .
In Figure 13 (right) this effect is demonstrated by presenting  
the derived \Htwo\ column density in the analytical model compared to the 
exact calculation for the 15 km s$^{-1}$ shock.   

The nature of analytical solution will change depending on the initial
conditions (shock velocity, pre-shock density).  Appropriate values
of  $n_p$ and $T_p$ for different solutions can be easily obtained from 
the ``plateau'' in  Figure 2.  However, {\em in practice only knowledge of
the local density and extinction/shielding need be identified (gas temperature
will track with extinction). With these few parameters one can reliably
solve for the formation of molecular hydrogen.}
In the case of CO its formation requires the
tracking of only dust shielding and could be computed via simple equilibrium calculations.

\begin{deluxetable}{lc}
\label{initabun}
\tablecolumns{2}
\tablewidth{2in}
\tablecaption{Initial Atomic Abundances\tablenotemark{a}}
\tablehead{
\colhead{Species} &
\colhead{log($\frac{N}{H}$) +12.00}}
\startdata
He & 10.93 \\
C  & 8.11 \\
N & 7.96  \\
O & 8.43 \\
Ne & 7.92 \\
Mg & 5.97 \\
Si & 6.29 \\
S & 7.20 \\
Ar & 6.42 \\
Ca & 2.60 \\
Fe & 5.33 \\
Ni & 3.56 \\
\enddata
\tablenotetext{a}{Estimated from Savage, Cardelli, \& Sofia (1992) and
Sofia, Cardelli, \& Savage (1994).}

\end{deluxetable}

\small
\begin{deluxetable}{lll}
\label{proc}
\tablecolumns{3}
\tablewidth{7in}
\tablecaption{Major Thermal and Chemical Processes}
\tablehead{
\colhead{Process} &
\colhead{Notes}
& \colhead{Reference}}
\startdata
\underline{Photoelectric heating}   &  &  \\
\hspace{0.2in}FUV field  &  Drain field; G$_0$ = 1.7 in units of Habing field  & (1) \\
\hspace{0.2in}Heating rate    &  Formalism of Bakes and Tielens (1994)   &  \\
\hspace{0.2in}Attenuation by dust  & Heating rate attenuated by e$^{-2.5{\rm A_V}}$; average attenuation  & (2) \\
                      & of UV field by dust between 1000 \AA\ -- 2000 \AA & \\
\underline{Cosmic Ray} && \\
\hspace{0.2in}H ionization rate  & $\zeta_{cr,\rm{H}} = 6 \times 10^{-18}$ s$^{-1}$ & (3) \\
\hspace{0.2in}H$_2$ ionization rate & $\zeta_{cr,\rm{H_2}} = 2\zeta_{cr,\rm{H}}$ & \\
\hspace{0.2in}Heating & $\Gamma_{cr} = 1 \times 10^{-27}n$  erg cm$^{-3}$ s$^{-1}$ & (4) \\
\underline{X-Ray} && \\
\hspace{0.2in}Ionization and heating & Expressions provided by Wolfire et al (1995); N$_{tot} = \int n_0v_0t$; & \\
                                         & n$_e$ from chemical equilibrium calculation   &  \\
\underline{Molecular Hydrogen} && \\
\hspace{0.2in}Formation rate &  $2.2 \times 10^{-18} S$ T$^{0.5}n$ s$^{-1}$; $S = 0.3$  & (5) \\
\hspace{0.2in}Formation heating & Formalism of Hollenbach \& McKee (1979); except only & \\
& 4\% of formation energy is deposited on grain surface & \\
& (Takahashi et al 2001) & \\
\hspace{0.2in}Self-shielding rate & Formalism of Draine \& Bertoldi (1996) & \\
\underline{Carbon Monoxide} && \\
\hspace{0.2in}Self-shielding rate & Formalism of Lee et al. (1996) & \\
\underline{Cooling} && \\
\hspace{0.2in}C II  & Impacts with e$^{-}$, H, H$_2$; $\gamma_{\rm H_2} = 0.5\gamma_{\rm H}$ & (6) \\
\hspace{0.3in}157.7 $\mu$m &\hspace{0.3in} A($^{2}P_{3/2}-^2P_{1/2}) = 2.29 \times 10^{-6}$ s$^{-1}$  & (7)\\
\hspace{0.2in}O I & Impacts with e$^{-}$, H, H$_2$, H$^+$; $\gamma_{\rm H_2} = 0.05\gamma_{\rm H}$ & (8) \\
\hspace{0.3in}145.6 $\mu$m &\hspace{0.3in} A($^{3}P_{0}-^3P_{1}) = 1.74 \times 10^{-5}$ s$^{-1}$ & (9) \\
\hspace{0.3in} 63.2 $\mu$m &\hspace{0.3in} A($^{3}P_{1}-^3P_{2}) = 8.96 \times 10^{-5}$ s$^{-1}$ & \\
\hspace{0.3in} 44.0 $\mu$m &\hspace{0.3in} A($^{3}P_{2}-^3P_{0}) = 1 \times 10^{-10}$ s$^{-1}$ & \\
\hspace{0.2in}C I & Impacts with H,H$_2$ &  (10)\\
\hspace{0.3in}610 $\mu$m &\hspace{0.3in} A($^{3}P_{1}-^3P_{0}) = 7.93 \times 10^{-8}$ s$^{-1}$ & (11) \\
\hspace{0.3in}370 $\mu$m &\hspace{0.3in} A($^{3}P_{2}-^3P_{1}) = 2.68 \times 10^{-7}$ s$^{-1}$ &  \\
\hspace{0.3in}980 $\mu$m &\hspace{0.3in} A($^{3}P_{2}-^3P_{0}) = 2 \times 10^{-14}$ s$^{-1}$ &  \\
\hspace{0.2in}H$^+$  & $3.5 \times 10^{-12}$(2/3)k(T/300)$^{-0.75}$Tn$_{H^+}$ erg s$^{-1}$ H$^{-1}$& \\
\hspace{0.2in}CO & $v = 0$ rotational transitions, collisions with H$_2$ & (12) \\
\hspace{0.2in}H$_2$ & not included due to low temperatures at time of formation & \\
\enddata
\tablecomments{\footnotesize
References: (1) Draine 1978; (2) Roberge et al 1991; (3) Millar, Farquhar, \& Willacy 
1997; (4) Goldsmith 2001; (5) Hollenbach, Werner, \& Salpeter 1971,
Leitch-Devlin \& Williams 1985; (6) Hayes \& Nussbaumer 1984, Launay \& Roueff 1977,
Hollenbach \& McKee 1979; (7) Nussbaumer \& Storey 1981; (8) Pequignot 1990,1996, Hollenbach \& McKee 1979;
(9) Baluja \& Zeippen 1988; (10) Launay \& Roueff 1977, Schr\"oder et al 1991; (11) Tielens
\& Hollenbach 1985; (12) Neufeld, Lepp, \& Melnick (1995), Neufeld \& Kaufman (1993)
}
\end{deluxetable}
\normalsize

\begin{center}
\begin{deluxetable}{llllll}
\label{optical}
\tablewidth{5.6in}
\tablecolumns{6}
\tablecaption{Optical Line Intensities (ergs~cm$^{-2}$~s$^{-1}$ sr$^{-1}$)}
\tablehead{
\colhead{Line} &
\colhead{10 $\rm km ^{-1}$} &
\colhead{15 $\rm km ^{-1}$} &
\colhead{20 $\rm km ^{-1}$} &
\colhead{30 $\rm km ^{-1}$} &
\colhead{50 $\rm km ^{-1}$}  
}
\startdata
H$\beta$    & $3.47\times 10^{-10}$& $4.68\times 10^{-10}$ & $6.38\times 10^{-10}$ & $5.76\times 10^{-9}$ & $7.19\times 10^{-8}$ \\
H$\alpha$   & $1.12\times 10^{-9}$& $1.45\times 10^{-9}$ & $1.44\times 10^{-8}$ & $2.94\times 10^{-8}$ & $2.48\times 10^{-7}$ \\
$[O I]$ 6300  & $3.85\times 10^{-10}$& $6.65\times 10^{-9}$ & $1.64\times 10^{-8}$ & $2.64\times 10^{-8}$ & $5.83\times 10^{-8}$ \\
$[S II]$ 6723 & $5.17\times 10^{-9}$& $1.47\times 10^{-8}$ & $4.77\times 10^{-8}$ & $7.16\times 10^{-8}$ & $2.15\times 10^{-7}$ \\
\enddata
\end{deluxetable}
\end{center}
 
\clearpage
\begin{center}
\begin{deluxetable}{lccccccccccccc}
\label{colden}
\tabletypesize{\small}
\rotate
\tablecolumns{14}
\tablewidth{8in}
\tablecaption{Column Densities for Atomic Species\tablenotemark{a}}
\tablehead{
\colhead{Model} &
\colhead{Hot H~I\tablenotemark{b}} &
\colhead{} &
\multicolumn{3}{c}{Cold H~I\tablenotemark{c}} &
\colhead{} &
\multicolumn{3}{c}{O~I} &
\colhead{} &
\multicolumn{3}{c}{C~II}  \\
\cline{2-2} \cline{4-6} \cline{8-10} \cline{12-14} \\ 
\colhead{v$_s$/n$_H$\tablenotemark{d}} &
\colhead{} &
\colhead{} &
\colhead{A$_v = 0.2$} &
\colhead{A$_v = 0.5$} &
\colhead{A$_v = 1.0$} &
\colhead{} &
\colhead{A$_v = 0.2$} &
\colhead{A$_v = 0.5$} &
\colhead{A$_v = 1.0$} &
\colhead{} &
\colhead{A$_v = 0.2$} &
\colhead{A$_v = 0.5$} &
\colhead{A$_v = 1.0$}  
}
\startdata
10/1 & 20.4 &&20.2 &20.8 &20.9 &&17.0 &17.4 &17.8 &&16.7 &17.1 &17.4 \\
10/2 & 19.7 &&20.5 &20.8 &20.9 &&17.0 &  17.4 &  17.8 &&  16.7&   17.1&   17.4 \\
10/3 &19.6  &&20.5 &20.8 &20.8 &&17.0 &  17.3 &  17.7 &&  16.7&   17.0&   17.3 \\
20/1 &19.8  &&20.2 &20.6 &20.6 &&17.0 &  17.5 &  17.7 &&  16.7&   17.0&   17.3 \\
30/1 &19.4  &&20.3 &20.4 &20.5 &&17.0 &  17.5 &  17.7 &&  16.7&   17.0&   17.3 \\
50/1 &18.9  &&20.0 &20.2 &20.2 &&17.0 &  17.3 &  17.6 &&  16.7&   17.0&   17.0 \\\\\hline\hline
\colhead{} &
\colhead{} &
\colhead{} &
\multicolumn{3}{c}{C~I}  & 
\colhead{} &
\multicolumn{3}{c}{CO}  \\
\cline{4-6} \cline{8-10}\\ 
\colhead{} &
\colhead{} &
\colhead{} &
\colhead{A$_v = 0.2$} &
\colhead{A$_v = 0.5$} &
\colhead{A$_v = 1.0$} &
\colhead{} &
\colhead{A$_v = 0.2$} &
\colhead{A$_v = 0.5$} &
\colhead{A$_v = 1.0$} \\\hline
10/1 &&& 13.9 &  15.0 &  15.9&& 8.5  &  12.5&   14.3 \\
10/2 &&&14.6  & 15.3  & 16.3 &&11.7  & 13.5 &  15.3 \\
10/3 &&&14.8  & 15.6  & 16.5 &&12.0  & 13.9 &  15.7  \\
20/1 &&&14.9  & 15.7  & 16.5 &&12.7  & 14.3 &  16.0 \\
30/1 &&&15.3  & 16.0  & 16.5 &&13.9  & 15.0 &  16.7 \\
50/1 &&&15.8  & 16.3  & 16.3 &&14.0  & 16.3 &  17.0 \\
\enddata
\tablenotetext{a}{Log$_{10}$ of column density in cm$^{-2}$ are given in the Table.}
\tablenotetext{b}{Column density of H~I with T $\gg$ 60 K.}
\tablenotetext{c}{Column density of H~I with T $<$ 60 K.}
\tablenotetext{d}{Units of $v_s$ is km s$^{-1}$ and $n$ is cm$^-3$.}
\end{deluxetable}
\end{center}

\begin{figure}
\includegraphics[height=16cm]{f10.eps}
\caption{Top panels show the time evolution of the H$_2$
(top left) and CO (top right) abundance.
Bottom panels show the evolution of the H$_2$ abundance 
(bottom left) and CO abundance (bottom right) as a function of
visual extinction.   All abundances are relative to total H. 
Each panel shows models with different initial H$_2$ fraction which are labeled
in the bottom left panel.  Each model adopts $v_0$ = 20 $\kms$  
and $n_0 = 1$ \cc .  The sharp rise in the H$_2$ abundance at t = 10$^6$ yr is
due to the non-inclusion of H$_2$ in the {\em Atomic} model.  Similarly the
rise in CO is due to formation from the existing H$_2$, but the abundance
drops because the molecules are unshielded to UV radiation. 
}
\label{co-h2}
\end{figure}

\begin{figure}
\includegraphics[height=11cm]{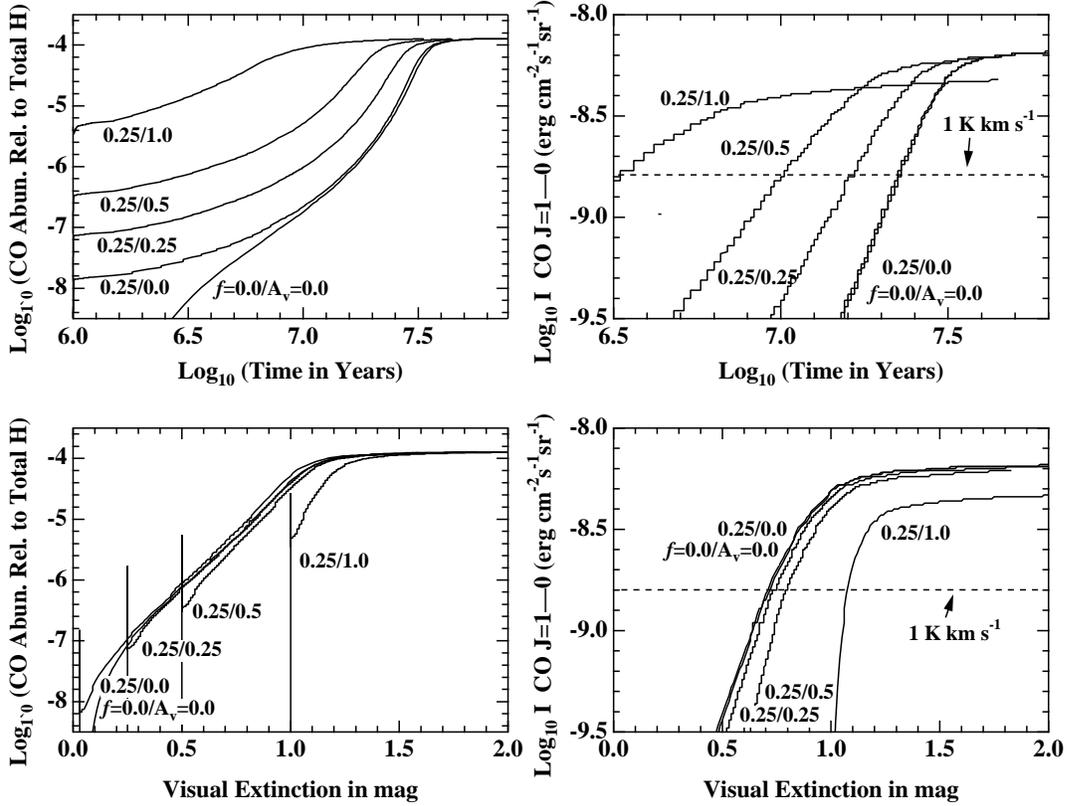}
\caption{
Left: Abundance of CO as a function of time for models with changing H$_2$
fraction and initial visual extinction in the pre-shock gas. 
Right:  
Predicted line intensities (erg cm$^{-2}$ s$^{-1}$ sr$^{-1}$) for CO J=1--0
arising from the plane-parallel slab
as a function of time for the same models as in the right-hand panel.
The dashed horizontal line denotes a CO J=1--0 integrated intensity of
1 K km s$^{-1}$.
}
\label{avinit}
\end{figure}

\begin{figure}
\includegraphics[height=8cm]{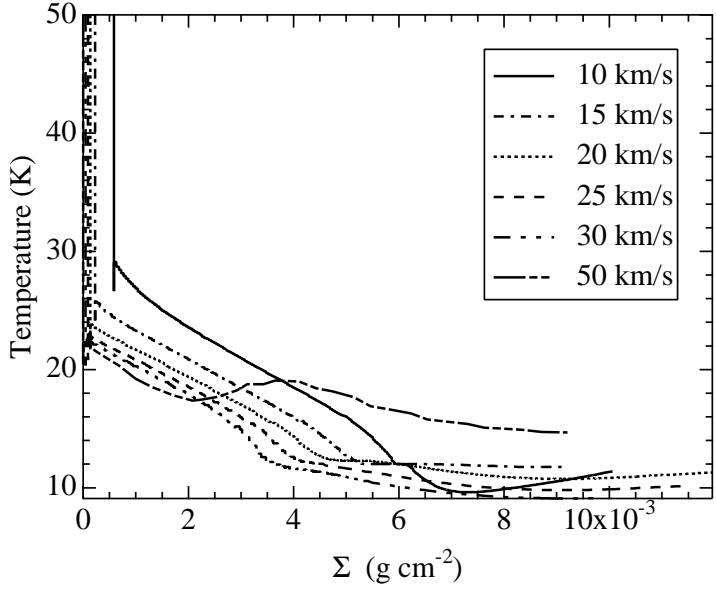}
\caption{
Plot showing the dependence of temperature with mass surface density 
(extinction) for solutions with variable shock speed and an initial
density of 1 cm$^{-3}$.
}
\label{temp-av}
\end{figure}

\begin{figure}
\plottwo{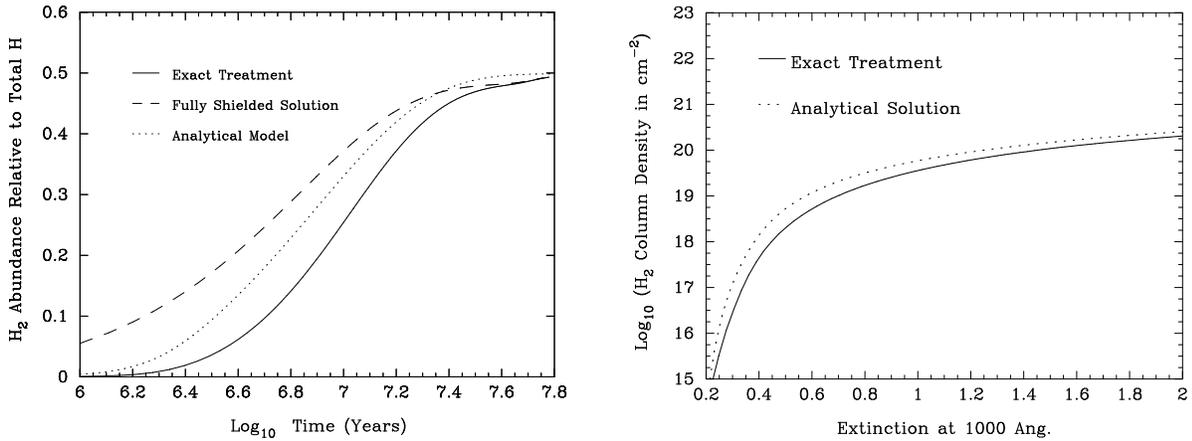}{f13b.eps}
\caption{Left:  Solid line shows exact treatment of the time evolution 
of the H$_2$ fraction for a 15 km s$^{-1}$ shock impinging on a medium
with a density of 1 cm$^{-3}$.  The dashed line shows the exact treatment
for the same model, only with the H$_2$ fully self-shielded at the start.
The dashed line presents the analytical approximation.  
Right: Comparison of the total \Htwo\ column density as a function of dust opacity
at 1000 \AA\ for the exact (solid) and analytical (dotted) solutions.
}
\end{figure}


\begin{thebibliography}{}

\bibitem[Aikawa \& Herbst(1999)]{ah99} Aikawa, Y.~\& Herbst, 
E.\ 1999, \aap, 351, 233 

\bibitem[Allen, Knapen, Bohlin, \& Stecher(1997)]{allen_h2} 
Allen, R.~J., Knapen, J.~H., Bohlin, R., \& Stecher, T.~P.\ 1997, \apj, 487, 171

\bibitem[Allen, Atherton, \& Tilanus(1986)]{allen_nat_h2} Allen, 
R.~J., Atherton, P.~D., \& Tilanus, R.~P.~J.\ 1986, \nat, 319, 296 171 

\bibitem[Arabadjis \& Bregman(1999)]{arabadjis_h2} Arabadjis, 
J.~S.~\& Bregman, J.~N.\ 1999, \apj, 510, 806 

\bibitem[Ballesteros-Paredes(2004)]{balles_cform}
Ballesteros-Paredes, J. 2004a. "Molecular Clouds. Formation and
Disruption", in "From Observations to Self-Consistent Modeling
of the Interstellar Medium", eds. M. Avillez \& D. Breitschwerdt.
Kluwer Academic Publishers (in press)

\bibitem[Ballesteros-Paredes, Hartmann, \&
V{\' a}zquez-Semadeni(1999)]{bhv99} Ballesteros-Paredes, J., Hartmann, L., \&
V\'azquez-Semadeni, E. 1999, \apj, 527, 285 (BHV)

\bibitem[Ballesteros-Paredes, V{\' a}zquez-Semadeni, \& 
Scalo(1999)]{bvs99} Ballesteros-Paredes, J., V{\' 
a}zquez-Semadeni, E., \& Scalo, J.\ 1999, \apj, 515, 286 


\bibitem[Baluja \& Zeippen(1988)]{1988JPhB...21.1455B} Baluja, K.~L.~\& 
Zeippen, C.~J.\ 1988, Journal of Physics B Atomic Molecular Physics, 21, 
1455 

\bibitem[Bergin, Neufeld, \& Melnick(1998)]{bmn98} Bergin, 
E.~A., Neufeld, D.~A., \& Melnick, G.~J.\ 1998, \apj, 499, 777 

\bibitem[Bergin \& Langer(1997)]{bl97} Bergin, E.~A.~\& 
Langer, W.~D.\ 1997, \apj, 486, 316 

\bibitem[Boulares \& Cox(1990)]{bc90}
Boulares, A., \& Cox, D.P. 1990, \apj, 365, 544

\bibitem[Brown, de Geus, \& de Zeeuw(1994)]{brown_ob1} Brown, A.G.A., de Geus, E.J., \& de Zeeuw, P.T. 1994,
\aap, 289, 101

\bibitem[Brown, Byrne, \& Hindmarsh(1989)]{vode} Brown, P.N., Byrne, G.D, \& 
Hindmarsh, A.C. 1989, SIAM J. Sci. Stat. Comput., 10 

\bibitem[Cazaux \& Tielens(2002)]{2002ApJ...575L..29C} Cazaux, S.~\& 
Tielens, A.~G.~G.~M.\ 2002, \apjl, 575, L29 

\bibitem[Cazaux \& Tielens(2004)]{2004ApJ...604..222} Cazaux, S.~\& 
Tielens, A.~G.~G.~M.\ 2004, \apj, 606, 222 

\bibitem[Cox(1972)]{cox72} Cox, D.~P.\ 1972, \apj, 178, 143 

\bibitem[Cox \& Raymond(1985)]{coxray85} Cox, D.~P.~\& Raymond, 
J.~C.\ 1985, \apj, 298, 651 


\bibitem[Dame(1993)]{dame93} Dame, T.\ M.\ 1993, AIP Conf.\ Proc.\ 278: Back to the
Galaxy, eds. S.S. Holt \& F. Verter, 267

\bibitem[Draine(1978)]{1978ApJS...36..595D} Draine, B.~T.\ 1978, \apjs, 36, 
595 

\bibitem[Draine \& Bertoldi(1996)]{draine_h2shield} Draine, B.~T.~\& 
Bertoldi, F.\ 1996, \apj, 468, 269 


\bibitem[Elmegreen(1991)]{elmegreen91} Elmegreen, B.\ G.\ 1991, NATO ASIC 342, The Physics of
Star Formation and Early Stellar Evolution, eds. C.J. Lada \&
N.J. Kylafis (Dordrecht: Kluwer), 35
%

\bibitem[Elmegreen(1993)]{elmegreen93} Elmegreen, B.G. 1993, in Protostars and Planets III,
ed. E.H. Levy \& J.I. Lunine, University of Arizona Press, Tucson, 97

\bibitem[Elmegreen \& Scalo(2004)]{elmegreen_araa} Elmegreen, B.G. \& Scalo, J. 2004, ARAA, 42, in press

\bibitem[Gibson, Taylor, Higgs, \& Dewdney(2000)]{gibson_hi} 
Gibson, S.~J., Taylor, A.~R., Higgs, L.~A., \& Dewdney, P.~E.\ 2000, \apj, 
540, 851 

\bibitem[Goldsmith(2001)]{2001ApJ...557..736G} Goldsmith, P.~F.\ 2001, 
\apj, 557, 736 

\bibitem[Haffner, Reynolds, \& Tufte(1999)]{1999ApJ...523..223H} Haffner, 
L.~M., Reynolds, R.~J., \& Tufte, S.~L.\ 1999, \apj, 523, 223 

\bibitem[Hartigan, Morse, \& Raymond(1994)]{1994ApJ...436..125H} Hartigan, 
P., Morse, J.~A., \& Raymond, J.\ 1994, \apj, 436, 125 

\bibitem[Hartmann, Ballesteros-Paredes, \& Bergin(2001)]{hbb01} Hartmann, L., Ballesteros-Paredes, J., \& Bergin, E.~A.\ 2001, 
\apj, 562, 852 

\bibitem[Hayes \& Nussbaumer(1984)]{1984A&A...134..193H} Hayes, M.~A.~\& 
Nussbaumer, H.\ 1984, \aap, 134, 193 

\bibitem[Heiles \& Troland(2003)]{ht03} Heiles, C.~\& 
Troland, T.~H.\ 2003, \apj, 586, 1067 

\bibitem[Hollenbach \& McKee(1979)]{hm79} Hollenbach, D.~\& 
McKee, C.~F.\ 1979, \apjs, 41, 555 

\bibitem[Hollenbach, Werner, \& Salpeter(1971)]{1971ApJ...163..165H} 
Hollenbach, D.~J., Werner, M.~W., \& Salpeter, E.~E.\ 1971, \apj, 163, 165 

\bibitem[Jenkins \& Tripp(2001)]{jenkins_tripp} Jenkins, E.~B.~\& 
Tripp, T.~M.\ 2001, \apjs, 137, 297 

\bibitem[Jura(1975)]{1975ApJ...197..575J} Jura, M.\ 1975, \apj, 197, 575 

\bibitem[Kaufman et al.(1999)]{kaufman_pdr} Kaufman, M.~J., Wolfire, M.~G., 
Hollenbach, D.~J., \& Luhman, M.~L.\ 1999, \apj, 527, 795 

\bibitem[Knee \& Brunt(2001)]{knee_brunt} Knee, L.~B.~G.~\& Brunt, 
C.~M.\ 2001, \nat, 412, 308 

\bibitem[Koyama \& Inutsuka(2000)]{koyama_cloudform} 
Koyama, H.~\& Inutsuka, S.\ 2000, \apj, 532, 980 

\bibitem[Kulkarni \& Heiles(1987)]{kulkarni_heiles} Kulkarni, S.~R.~\& 
Heiles, C.\ 1987, ASSL Vol.~134: Interstellar Processes, 87 

\bibitem[Launay \& Roueff(1977)]{1977A&A....56..289L} Launay, J.~M.~\& 
Roueff, E.\ 1977, \aap, 56, 289 

\bibitem[Lee et al.(1996)]{lee_coshield} Lee, H.-H., Herbst, E., 
Pineau des Forets, G., Roueff, E., \& Le Bourlot, J.\ 1996, \aap, 311, 690 

\bibitem[Leitch-Devlin \& Williams(1985)]{1985MNRAS.213..295L} 
Leitch-Devlin, M.~A.~\& Williams, D.~A.\ 1985, \mnras, 213, 295 

\bibitem[Li \& Goldsmith(2003)]{li_hi} Li, D.~\& Goldsmith, 
P.~F.\ 2003, \apj, 585, 823 

\bibitem[Mac Low \& Klessen(2004)]{maclow_rvmp} Mac Low, M.~\& 
Klessen, R.~S.\ 2004, Reviews of Modern Physics, 76, 125 

\bibitem[Mac Low et al.(1998)]{maclow98} Mac Low, M.-M., Klessen, R. S., Burkert, A., \& Smith,
M. D. 1998, Phys. Rev. Lett., 80, 275


\bibitem[McCray \& Kafatos(1987]{mk87} McCray, R.~\& Kafatos, M.\ 1987, \apj, 317, 190 
 
\bibitem[Millar, Farquhar, \& Willacy(1997)]{umist} Millar, 
T. J., Farquhar, P. R. A. \& Willacy, K. 1997, \aaps, 121, 139 

\bibitem[Minter \& Spangler(1997)]{1997ApJ...485..182M} Minter, A.~H.~\& 
Spangler, S.~R.\ 1997, \apj, 485, 182 

\bibitem[Nagai, Inutsuka, \& Miyama(1998)]{nim98} 
Nagai, T., Inutsuka, S., \& Miyama, S.M. 1998, \apj, 506, 306

\bibitem[Neufeld, Lepp, \& Melnick(1995)]{1995ApJS..100..132N} Neufeld, 
D.~A., Lepp, S., \& Melnick, G.~J.\ 1995, \apjs, 100, 132 

\bibitem[Neufeld \& Kaufman(1993)]{1993ApJ...418..263N} Neufeld, D.~A.~\& 
Kaufman, M.~J.\ 1993, \apj, 418, 263 

\bibitem[Neufeld \& Dalgarno(1989)]{nd89} Neufeld, D.~A.~\& 
Dalgarno, A.\ 1989, \apj, 340, 869 

\bibitem[Nussbaumer \& Storey(1981)]{1981A&A....96...91N} Nussbaumer, H.~\& 
Storey, P.~J.\ 1981, \aap, 96, 91 
%
\bibitem[Ostriker, Stone, \& Gammie(2001)]{osg} Ostriker, 
E.~C., Stone, J.~M., \& Gammie, C.~F.\ 2001, \apj, 546, 980 

\bibitem[Padoan \& Nordlund(1999)]{1999ApJ...526..279P} Padoan, P.~\& 
Nordlund, {\AA}.\ 1999, \apj, 526, 279 

\bibitem[Padoan(1995)]{1995MNRAS.277..377P} Padoan, P.\ 1995, \mnras, 277, 
377

\bibitem[Passot \& V{\' a}zquez-Semadeni(1998)]{pv98} 
Passot, T.~\& V{\' a}zquez-Semadeni, E.\ 1998, \pre, 58, 4501 

\bibitem[Pavlovski, Smith, Mac Low, \& Rosen(2002)]{pavlovski02} 
Pavlovski, G., Smith, M.~D., Mac Low, M., \& Rosen, A.\ 2002, \mnras, 337, 
477 

\bibitem[Pequignot(1996)]{1996A&A...313.1026P} Pequignot, D.\ 1996, \aap, 
313, 1026 

\bibitem[Pequignot(1990)]{1990A&A...231..499P} Pequignot, D.\ 1990, \aap, 
231, 499 

\bibitem[Pringle, Allen, \& Lubow(2001)]{pringle_h2} Pringle, 
J.~E., Allen, R.~J., \& Lubow, S.~H.\ 2001, \mnras, 327, 663 


\bibitem[Reach et al.(1995)]{1995ApJ...451..188R} Reach, W.~T.~et al.\ 
1995, \apj, 451, 188 

\bibitem[Raymond(1979)]{raymond79} Raymond, J.~C.\ 1979, \apjs, 
39, 1 

\bibitem[Roberge, Jones, Lepp, \& Dalgarno(1991)]{1991ApJS...77..287R} 
Roberge, W.~G., Jones, D., Lepp, S., \& Dalgarno, A.\ 1991, \apjs, 77, 287 

\bibitem[Roberts \& Stewart(1987)]{roberts_spiral} Roberts, W.~W.~\& 
Stewart, G.~R.\ 1987, \apj, 314, 10 

\bibitem[Sasao(1973)]{sasao} Sasao, T.\ 1973, \pasj, 25, 1 


\bibitem[Savage, Cardelli, \& Sofia(1992)]{savage_zetaoph} Savage, 
B.~D., Cardelli, J.~A., \& Sofia, U.~J.\ 1992, \apj, 401, 706 

\bibitem[Savage et al.(1977)]{savage_h2} 
Savage, B.~D., Drake, J.~F., Budich, W., \& Bohlin, R.~C.\ 1977, \apj, 216, 
291 


\bibitem[Schneider \& Elmegreen(1979)]{se79} Schneider, S.~\& Elmegreen, B.~G.\ 1979, \apjs, 41, 87

\bibitem[Schroder et al.(1991)]{1991JPhB...24.2487S} Schroder, K., 
Staemmler, V., Smith, M.~D., Flower, D.~R., \& Jaquet, R.\ 1991, Journal of 
Physics B Atomic Molecular Physics, 24, 2487

\bibitem[Shull(1993)]{shull93} Shull, J.M. 1993, in Massive Stars: Their Lives in the Interstellar
Medium, eds. J.P. Cassinelli \& E.B. Churchwell, ASP Conf. Ser. 35, 327

\bibitem[Shu et al.(1972)]{shu_spiral} Shu, F.~H., Milione, V., 
Gebel, W., Yuan, C., Goldsmith, D.~W., \& Roberts, W.~W.\ 1972, \apj, 173, 
557 

\bibitem[Shull et al.(2000)]{shull_fuse} Shull, J.~M.~et al.\ 
2000, \apjl, 538, L73 

\bibitem[Sofia, Cardelli, \& Savage(1994)]{sophia_dust} Sofia, 
U.~J., Cardelli, J.~A., \& Savage, B.~D.\ 1994, \apj, 430, 650 

\bibitem[Spitzer(1978)]{spitzer78} Spitzer, L.\ 1978, New York 
Wiley-Interscience, 1978

\bibitem[Stone et al.(1998)]{sog} Stone, J. M., Ostriker, E. C., \& Gammie, C. F. 1998,
ApJ, 508, L99

\bibitem[Tielens \& Hollenbach(1985)]{1985ApJ...291..722T} Tielens, 
A.~G.~G.~M.~\& Hollenbach, D.\ 1985, \apj, 291, 722 
%
\bibitem[Tilanus \& Allen(1989)]{tilanus_h2} Tilanus, R.~P.~J.~\& 
Allen, R.~J.\ 1989, \apjl, 339, L57 
%
\bibitem[V\'azquez-Semadeni(1994)]{1994ApJ...423..681V} V\'azquez-Semadeni, E.\ 
1994, \apj, 423, 681 

\bibitem[Vishniac(1994)]{vishniac94} Vishniac, E. 1994, \apj, 428, 186
%
\bibitem[VonWeiczacker(1951)]{vonweiczacker} VonWeiczacker, C. F. 1951. \apj,
114, 165

\bibitem[Weaver et al.(1977)]{weaver77} Weaver, R., McCray, R., 
Castor, J., Shapiro, P., \& Moore, R.\ 1977, \apj, 218, 377 
%
\bibitem[Wolfire et al.(1995)]{wolfire95} Wolfire, M.~G., 
Hollenbach, D., McKee, C.~F., Tielens, A.~G.~G.~M., \& Bakes, E.~L.~O.\ 
1995, \apj, 443, 152 
%
%
%
%
\end{thebibliography}
\end{document}